\documentclass[10pt,doublecolumn]{IEEEtran}
\usepackage{color}
\usepackage{float}
\usepackage{url}
\usepackage{amsmath}
\usepackage{amssymb}
\usepackage{amsfonts}
\usepackage{epsfig}
\usepackage{graphicx}
\usepackage{theorem}
\usepackage{cite}
\usepackage{euscript}
\usepackage{pifont}
\usepackage{fancyhdr}
\usepackage{subfigure}
\usepackage{setspace}
\usepackage{hyperref}
\usepackage{verbatim}
\usepackage{multirow}
\usepackage{array}
\usepackage{flushend}
\usepackage{algorithmic}
\usepackage{algorithm}

\newcommand{\mb}{\mathbf}

\newcommand{\ew}{\textrm{ew}}
\newtheorem{definition}{Definition}
\newtheorem{lemma}{Lemma}
\newtheorem{proposition}{Proposition}
\newtheorem{corollary}{Corollary}

\renewcommand{\b}{\textbf}
\newcommand{\0}{\b 0}
\newcommand{\I}{\b I}
\newcommand{\T}{\mathcal{T}}
\newcommand{\R}{\mathcal{R}}
\newcommand{\Set}{\mathcal{S}}
\newcommand{\lf}{\left\lfloor}
\newcommand{\rf}{\right\rfloor}
\newcommand{\lc}{\left\lceil}
\newcommand{\rc}{\right\rceil}
\newcommand{\GLNC}{\textrm{GLNC}}
\newcommand{\mmax}{{\textrm{max}}}

\newcommand{\Prob}{\textrm{Pr}}

\newcolumntype{x}[1]{%
>{\centering\hspace{0pt}}p{#1}}%
\newcommand{\tn}{\tabularnewline}

\newcommand{\ind}{~1\hspace{-3mm}{1}~}

\begin{document}

\hyphenation{time-slots}

\title{Diversity Analysis, Code Design and Tight Error Rate Lower Bound for Binary Joint Network-Channel Coding}
\author{Dieter~Duyck,~Michael~Heindlmaier,~Daniele~Capirone,~and~Marc~Moeneclaey
\thanks{\noindent This work was supported by the European Commission in the framework of the FP7 Network of Excellence in Wireless COMmunications NEWCOM++ (contract n. 216715).}%
\thanks{Dieter  Duyck  and  Marc Moeneclaey  are  with  the Department  of Telecommunications  and  Information processing,  Ghent
University,   St-Pietersnieuwstraat    41,   B-9000   Gent,   Belgium, \{dduyck,mm\}@telin.ugent.be.}%
\thanks{Michael Heindlmaier is with Technische Universit{\"a}t M{\"u}nchen in M{\"u}nchen, Germany, michael.heindlmaier@tum.de.}%
\thanks{Daniele Capirone was with Politecnico di Torino in Torino, Italy, capirone.polito@gmail.com.}}


\maketitle

\begin{abstract}
Joint network-channel codes (JNCC) can improve the performance of communication in wireless networks, by combining, at the physical layer, the channel codes and the network code as an overall error-correcting code. JNCC is increasingly proposed as an alternative to a standard layered construction, such as the OSI-model. The main performance metrics for JNCCs are scalability to larger networks and error rate. The diversity order is one of the most important parameters determining the error rate. 
The literature on JNCC is growing, but a rigorous diversity analysis is lacking, mainly because of the many degrees of freedom in wireless networks, which makes it very hard to prove general statements on the diversity order.
In this paper, we consider a network with slowly varying fading point-to-point links, where all sources also act as relay and additional non-source relays may be present. We propose a general structure for JNCCs to be applied in such network. In the relay phase, each relay transmits a linear transform of a set of source codewords. 
Our main contributions are the proposition of an upper and lower bound on the diversity order, a scalable code design and a new lower bound on the word error rate to asses the performance of the network code. The lower bound on the diversity order is only valid for JNCCs where the relays transform only two source codewords. We then validate this analysis with an example which compares the JNCC performance to that of a standard layered construction.
Our numerical results suggest that as networks grow, it is difficult to perform significantly better than a standard layered construction, both on a fundamental level, expressed by the outage probability, as on a practical level, expressed by the word error rate.
\end{abstract}

\section{Introduction}
\label{sec: introduction}

Point-to-point communication has revealed many of its secrets. Driven by new applications, research in wireless communication is now focusing more on the optimization of communication in wireless \textit{networks}. For example, the joint operation of multiple network layers can be optimized, denoted as cross-layer design \cite{shakkottai2003cld, srivastava2005cld}, thereby leaving the classical layered architectures, such as the seven-layer open systems interconnect (OSI) model \cite[p. 20]{bertsekas1992dn}. Another example of network optimization is cooperative communication, where multiple nodes in the network cooperate to improve their error performance. Cooperation may occur in many forms at different layers, e.g. cooperative channel coding at the physical layer and network coding at the network layer. Network coding refers to the case where the intermediate nodes in the network are allowed to perform encoding operations over multiple received streams from different sources. In a standard layered construction, the decoding of the network code is performed at the network layer, after the point-to-point transmissions have been decoded at the physical layer. Channel coding refers to the case where nodes perform coding over \textit{one} point-to-point wireless link only. Cooperative channel coding is achieved by letting one or more relays transmit redundant bits for \textit{one} source at a time. Usually, channel coding and network coding are studied separately (e.g. \cite{laneman2004cdw, hunter2004cc, duy2009ldg} for cooperative channel coding and \cite{ahl2000nif, li2003lnc, koetter2003algebraic, rebellato2010muc, xiao2009muc} for network coding). 

\textit{Joint network-channel coding} (JNCC) received much of attention in the last years. This technique combines both decode and forward \cite{kramer2005csa} (cooperative communication) and cross-layer design by using a network code, which is accessible at the physical layer. The rationale behind joint network-channel coding is to improve the performance, by combining, at the physical layer, the channel codes and the network code as an overall error-correcting code \cite{guo2009apj}. Mostly, the two most important performance metrics are $(R, P_e)$, where $R$ is the spectral efficiency and $P_e$ is the error rate (bit error rate or word error rate). Here, we consider a fixed spectral efficiency $R$, so that the aim is to minimize $P_e$ for a given channel quality, expressed by $\gamma$, the signal-to-noise ratio (SNR). Expressing the asymptotic (for large $\gamma$) error rate as $P_e = \frac{1}{c \gamma^d}$, where $c$ and $d$ are defined as the \textit{coding gain} and the \textit{diversity order}, respectively, improving the performance refers to maximizing first $d$ and then $c$ (because $d$ has the larger impact).

Standard linear network coding consists of taking \textit{linear combinations} of several source packets, well known as the typical XOR operations for binary codes. In general, non-binary coefficients are used in the linear combinations. However, when the network code is used at the physical layer to decode the noisy channel output, this simple technique might yield poor error performance. Therefore, powerful network codes consisting of taking \textit{linear transformations} of the incoming information packets, have been introduced. We denote this methodology as generalized linear network coding (GLNC). The well known standard linear network codes, taking \textit{linear combinations}, are a special case of GLNC. Combining GLNC with channel coding, is denoted as joint network-channel coding (JNCC). The JNCC, which is the overall code comprising the channel codes and the network code, can for example be an LDPC code or a Turbo code. Of course, while JNCC brings more degrees of freedom and opens perspectives for a higher coding gain $c$, it must be verified that important metrics, such as the diversity order $d$ and the scalability to large networks, are not negatively affected.

Binary JNCC has already been studied in the literature. Pioneering papers \cite{hausl2006jnc, hausl2005ina} designed Turbo codes and LDPC codes, respectively, for the multiple access relay channel\footnote{In the MARC, two sources are helped by one relay to communicate to the destination.} (MARC) and for the two-way relay channel \cite{hausl2006ina}. However, the code design was not immediately scalable to general large networks and did not contain the required structure to achieve full diversity. In \cite{duyck2009afj, duyck2010aac}, a full-diversity JNCC for the MARC was proposed but it was not extended to large networks. The work of Hausl et al. \cite{hausl2006jnc, hausl2005ina, hausl2006ina} was followed by the interesting work of Bao et al. \cite{Bao2007gan}, presenting a JNCC that is scalable to large networks. However, this JNCC was not structured to achieve full diversity and has weak points from a coding point of view \cite{duyck2011tfd}. A deficiency in the literature, for general networks with sources and relays, is the lack of a detailed diversity analysis in the case that the sources can act as a relay (which is for example the model assumed by \cite{Bao2007gan}). The effect of the parameters of the JNCC on the diversity order is in general not known, because it is very hard to prove general statements on the diversity order, in an environment with so many degrees of freedom. This paper is a modest attempt to contribute to the solution of this problem. Related to this, we mention \cite{Li2011ncl, Li2011bfn}, where the authors designed a JNCC for the 
case where the sources cannot act as a relay, but other nodes play the role of relay to communicate to one destination. As the source nodes are excluded to act as a relay node in this model, the diversity analysis in \cite{Li2011ncl, Li2011bfn} is different from ours.

In this paper, we consider a JNCC where the network code forms an integral part of the overall error-correcting code, that is used at the destination to decode the information from the sources. The body of this paper consists of two main parts. In Sec. \ref{sec: diversity analysis of JNCC}, we perform a diversity analysis, leading to an upper bound on the diversity order of any linear binary JNCC following our system model, and to a lower bound on the diversity order for a particular subset of linear binary JNCCs. The upper and lower bound depend on the parameters of the JNCC and can be used to verify whether a particular JNCC has the potential to achieve full diversity on a certain network. Secondly, in Sec. \ref{sec: practical JNCC for n_{u_r}=2}, a specific JNCC of the LDPC-type is proposed that achieves full diversity for a well identified set of wireless networks. The scalability of this specific JNCC to large networks is discussed. The coding gain $c$ is not considered in the body of the paper and the parameters of our proposed code may be further optimized by applying techniques such as in \cite{Li2011ncl}, to maximize $c$. To assess the performance of the proposed JNCC, we determine the outage probability, a well known lower bound of the word error rate, in Sec. \ref{sec: lower bound on the WER}. We also present a tighter word error rate lower bound in Sec. \ref{sec: Calculation of tight lower bound on WER}, that takes into account the particular structure of the JNCC. In Sec. \ref{sec: numerical results}, the numerical results corroborate the established theory. We also briefly comment on the coding gain achieved by the proposed JNCC and conclusions are drawn for different classes of large networks. 

This paper extends the work, published in \cite{duyck2011tfd}, by also considering non-perfect source-relay channels, by considerably extending the diversity analysis, by providing an achievability proof for the diversity order of the proposed JNCC, by clearly indicating the set of wireless networks where the proposed JNCC is diversity-optimal, by providing a tighter lower bound on the word error rate, and by providing more numerical results. 

\section{Joint network-channel coding}

We first illustrate joint network-channel coding by means of a simple example. Consider two sources orthogonally broadcasting a vector of symbols, mapped from the binary vectors $\mb{s}_1$ and $\mb{s}_2$, respectively, to a relay and a destination. This channel is denoted as a multiple access relay channel (MARC) in the literature. Supposing that the relay is able to decode the received symbols, the relay computes a binary vector $\mb{r}_1$, which is mapped to symbols and transmitted to the destination. The relation between all bits is expressed by the JNCC, whose parity-check matrix has the following general form,
\begin{equation}
\label{eq: example JNCC}
H = 
\scriptsize
\stackrel{
\left.
\begin{array}{x{0.6cm}x{0.6cm}x{0.6cm}}
$\mb{s}_1$ & $\mb{s}_2$ & $\mb{r}_1$
\end{array}
\right.
}{
\left[ \begin{array}{x{0.6cm}x{0.6cm}x{0.6cm}}
$H_{p}$ & \0 & \0 \tn
\0 & $H_{p}$ & \0  \tn
\0 & \0 & $H_{p}$  \tn
$H_{1}^1$ & $H_{2}^1$ & $H_{1}$ 
\end{array}  \right]}
\normalsize .
\end{equation}
The matrix $H_{p}$ represents the parity-check matrix for the point-to-point channel code. Each of the binary vectors $\mb{s}_1$, $\mb{s}_2$ and $\mb{r}_1$, can be separately decoded using this code. The bottom part of $H$ represents the GLNC, which we denote as $H_{\GLNC} = [H_{1}^1~ H_{2}^1~H_{1}]$. It expresses the relation between $\mb{r}_1$, $\mb{s}_1$ and $\mb{s}_2$. More specifically, 
we have
\begin{equation}
\label{eq: MARC equation}
	H_{1}\mb{r}_1=H_{1}^1\mb{s}_1+H_{2}^1\mb{s}_2.
\end{equation}	
Note that GLNC includes standard network codes used in an OSI communication model as a special case. In the latter case, the matrices $H_{j}^i$ and $H_{i}$ (considering more than one relay in general) are identity matrices or all-zero matrices, so that the network code simplifies to the relay packet being a linear combination of source packets, also expressed as XORing of packets or symbol-wise addition of packets.

Ideally, the overall matrix $H$ conforms optimized degree distributions that specify the LDPC code. When the channels between sources and relay are perfect, we can drop the first 3 sets of rows and only keep the GLNC, represented by $H_{\GLNC}$; in this case the information bits of the code are $\mb{s}_1$ and $\mb{s}_2$, and $\mb{r}_1$ contains the parity bits. This is still a JNCC as the redundancy in the network code is used to decode the received symbols on the physical layer at the destination. In \cite{duyck2010aac, duyck2009afj}, it is proved that the matrices $H_{p}$ do no affect the diversity order in the case of the MARC.

\section{System model}
\label{sec: system model}

We consider wireless networks with $m_s$ sources directly communicating to a common destination (e.g. cellphones communicating to a base station). Two time-orthogonal phases are distinguished. In the \textit{source phase}, the sources orthogonally broadcast their respective source packet. In the following \textit{relay phase}, the relays orthogonally broadcast their respective packet. All considered sources overhear each other during the source phase, and act as relay in the relay phase. Other nodes, not acting as a source, might be present in the network (i.e., overhearing the sources) and also act as relay. Hence, we consider a total of $m_r$ relays, where $m_r \geq m_s$. This general network model, which is practically relevant as it fits many applications, is adopted in e.g. \cite{Bao2007gan}. Take for example any large network and consider a volume in space (cfr. picocells or femtocells) where all nodes can overhear each other. These nodes form \textit{sub-networks} and can be modelled by our proposed model. Note that in the literature, sometimes other models are assumed, such as the $M-N-1$ model \cite{Li2011ncl, Li2011bfn}, where $M$ sources are helped by $N$ relays (the relays are nodes different from the sources) to communicate to one destination. 

All devices have one antenna, are half-duplex and transmit orthogonally using BPSK modulation. The $K$ information bits of each source are encoded via \textit{point-to-point channel codes} into a systematic codeword, denoted as \textit{source codeword}, of length $L$, expressed by the column vector $\mb{s}_{u_s}$ for user ${u_s}$, ${u_s} \in [1, \ldots, m_s]$. The parity-check matrix of dimension $(L-K) \times L$ of this point-to-point codeword is denoted by $H_p$, which is the same for each user ${u_s}$, so that $H_p \mb{s}_{u_s} = \mb{0}$ for all ${u_s}$. In the relay phase, each relay ${u_r}$, ${u_r} \in [1, \ldots, m_r]$, transmits a point-to-point codeword $\mb{r}_{u_r}$ of length $L$ to the destination, also satisfying $H_p \mb{r}_{u_r} = \mb{0}$. Hence, all slots have equal duration, the coding rate of the point-to-point channels is $R_{c,p} = \frac{K}{L}$, and the overall coding rate is $R_c = \frac{m_s K}{(m_s+m_r)L} = R_{c,p} \frac{m_s}{m_s+m_r}$. We define the fraction of source transmissions in the total number of transmissions as the network coding rate $R_n=\frac{m_s}{m_s+m_r}$, so that $R_c = R_{c,p} R_n$. The overall codeword of length $(m_s+m_r)L$ is expressed by the column vector 
\begin{equation}
\label{eq: overall codeword}
	\mb{x} = [\mb{s}_1^T \ldots \mb{s}_{m_s}^T \mb{r}_1^T \ldots \mb{r}_{m_s}^T \ldots \mb{r}_{m_r}^T]^T.
\end{equation}
The destination declares a word error if it can not perfectly retrieve all $m_s K$ information bits, and the overall word error rate is denoted by $P_{\ew}$.

All relevant channels between different\footnote{Unless mentioned otherwise, we assume that channels are reciprocal, i.e., the channel from $u_1$ to $u_2$ is the same as the channel from $u_2$ to $u_1$.} pairs of network nodes are assumed independent, memoryless, with real additive white Gaussian noise and multiplicative real fading (Rayleigh distributed with expected squared value equal to one). The fading coefficient of a wireless link is only known at the receiver side of that link. We consider a slow fading environment with a finite coherence time that is longer than the duration of the source phase \textit{and} the relay phase, so that the fading gain between two network nodes takes the same value during both phases. We denote the fading gain from node $u$ to the destination as $\alpha_u$, with $\mathbb{E}[\alpha_u^2]=1$. All point-to-point channels have the same average signal-to-noise ratio (SNR), denoted by $\gamma$. Differences in average SNR between the channels would not alter the diversity analysis, on the condition that the large SNR behaviour inherent to a diversity analysis refers to all\footnote{In practice, increasing the SNR value can be achieved by increasing the \textit{transmission} power of a node, so that both the SNR of the node-to-destination channels and channels between non-destination nodes increase.} SNRs being large. Denoting the received symbol vector at the destination\footnote{For conciseness, we do not formulate the equation for channels between non-destination nodes.} in timeslot $i$ as $\mb{y}_i$, the channel equation is
\begin{equation}
\label{eq: channel equation}
\left\{
	\begin{array}{l}
	\mb{y}_{u_s} = \alpha_{u_s} \mb{s}_{u_s}^\prime + \mb{n}_{u_s},~~{u_s} =1, \ldots, m_s \\
	\mb{y}_{m_s+u_r} = \alpha_{u_r} \mb{r}_{u_r}^\prime + \mb{n}_{m_s+{u_r}},~~{u_r} =1, \ldots, m_r ,
	\end{array}
	\right.
\end{equation}
where $\mb{n}_i \sim \mathcal{CN}(\mb{0}, \frac{1}{\gamma} I)$ is the noise vector in timeslot $i$, $\mb{s}_{u_s}^\prime = 2 \mb{s}_{u_s} - 1$ and $\mb{r}_{u_r}^\prime = 2 \mb{r}_{u_r} - 1$ (BPSK modulation).

Hence, at the destination, each of the $m_s$ independent fading gains between the sources and the destination affects $2 L$ bits ($L$ bits in the source phase and $L$ bits in the relay phase) and each of $m_r-m_s$ fading gains between the non-source relays and the destination affects $L$ bits, assuming that all $m_r$ relays could decode the messages received from the sources. Hence, from the point of view of the destination, the overall codeword is transmitted on a block fading (BF) channel with $m_r$ blocks, each affected by its own fading gain, where $m_s$ blocks have length $2L$ and $m_r-m_s$ blocks have length $L$. This notion will be essential in the subsequent diversity analysis (Sec. \ref{sec: diversity analysis of JNCC}).

In the source phase, relay ${u_r}$ attempts to decode the received symbols from sources belonging to the decoding set $\Set({u_r})$. The users that are successfully decoded at relay ${u_r}$ are added to its retrieval set, denoted by $\R({u_r})$, $\R({u_r}) \subset \Set({u_r})$, with cardinality $l_{u_r}$. Next, in the relay phase, relay ${u_r}$ transmits a relay packet, which is a linear transformation of $n_{u_r}$ source codewords\footnote{Note that relays $u$ are not allowed to consider relay codewords $\mb{r}_{u_r}$ for inclusion in $\Set({u})$. As a consequence, the right part of $H_{\GLNC}$ is diagonal in Eq. \ref{eq: H_GLNC}. This restriction was not always applied in the literature (e.g., \cite{Bao2007gan}), but it simplifies the theoretical analysis and code design.} originated by the sources from the transmission set $\T({u_r})=\{u_1,~ \ldots,~ u_{n_{u_r}}\}$ of relay $u_r$, with $\T({u_r}) \subset \R({u_r})$. If $l_{u_r} < n_{u_r}$, then relay ${u_r}$ does not transmit anything. In Sec. \ref{sec: diversity analysis of JNCC}, we show that $n_{u_r}$ is an important parameter that strongly affects the diversity order.

For example, user $3$ attempts to decode the messages from users $1$, $2$ and $5$, and succeeds in decoding the messages from users $1$ and $5$ from which a linear transformation is computed. Hence, $\Set(3)=\{1,~2,~5\}$, $\R(3)=\T(3)=\{1,~5\}$, $l_3=n_3 = 2$. Because the channel between a node and the destination remains constant during both source and relay phases, a relay has no interest in including its own source message in $\Set({u_r})$.

Using the transmission set for each relay, the GLNC in Eq. (\ref{eq: MARC equation}) generalizes to 
\begin{equation}
\label{eq: relay}
 H_{u_r} \mb{r}_{u_r} = \bigoplus_{{u_s} \in \mathcal{T}({u_r})} H_{u_s}^{u_r} \mb{s}_{u_s},
\end{equation}
where the matrices $H_{u_r}$ and $H_{u_s}^{u_r}$ are of dimension $K \times L$. Hence, each transmitted relay codeword $\mb{r}_{u_r}$ is a linear transformation of $n_{u_r}$ source codewords. The superscript ${u_r}$ in $H_{u_s}^{u_r}$ indicates that the vector $\mb{s}_{u_s}$ is in general not transformed by the same matrix for all relays ${u_r}$ where ${u_s} \in \mathcal{T}({u_r})$. The overall parity-check matrix $H$ is thus expressed as 

\begin{equation}
\label{eq: general parity-check matrix form}
 H =  \left[ \begin{array}{c}
             H_c \\
	     	H_{\GLNC}\\
            \end{array}\right] ,
\end{equation}
where $H_c$ is block diagonal with $H_p$ on its diagonal, representing the channel code, and
\begin{equation}
\label{eq: H_GLNC}
\textstyle
 H_{\GLNC} = \left[ \begin{array}{ccccccc}
             H_{1}^1 & \ldots & H_{m_s}^1 & H_{1} & 0 & \ldots & 0 \\
	     H_{1}^2 & \ldots & H_{m_s}^2 & 0 & H_{2} & \ldots & 0 \\
	    \vdots & \vdots &\vdots &\vdots &\vdots &\ddots &\vdots \\
	     H_{1}^{m_r} & \ldots & H_{m_s}^{m_r} & 0 & 0 & \ldots & H_{m_r} \\
            \end{array}\right]
\end{equation}
represents the GLNC.

\section{Diversity analysis of JNCC}
\label{sec: diversity analysis of JNCC}

Before passing to the actual diversity analysis, we provide the well-known formal definition of the diversity order \cite{tse2005fwc}.
\begin{definition}
\label{def: div block fading}
The diversity order attained by a code $\mathcal{C}$ is defined as 
\begin{displaymath} 
	d = -\lim_{\gamma \rightarrow \infty} \frac{\log P_{\ew}}{\log \gamma}. 
\end{displaymath} 
\end{definition}
In other words, $P_{\ew} \propto \gamma^{-d}$, where $\propto$ denotes \textit{proportional to}.

In the proofs of propositions in this paper, we will often use the diversity equivalence between a BF channel and a block binary erasure channel (block BEC), which was proved in \cite{bou2011dac, bou2012tde}. A block BEC channel is obtained by restricting the fading gains in our model to belong to the set $\{0, \infty\}$, so that a point-to-point channel is either erased or perfect. Denoting the erasure probability $\Prob[\alpha_{u_r}=0]$ by $\epsilon$, a diversity order $d$ is achieved if $P_{\ew} \propto \epsilon^d$ for small $\epsilon$ \cite{fabregas2006cit}. A diversity order of $d$ is thus achievable if there exists no combination of $d-1$ erased point-to-point channels leading to a word error. On the other hand, a diversity order of $d$ is not achievable if there exists at least one combination of $d-1$ erased channels leading to word error.

In this section, we present the relation between the diversity order $d$ and the parameters $\{n_{u_r}, u_r=1, \ldots, m_r\}$, as well as between $d$ and the choice of $\{\T({u_r}), {u_r}=1, \ldots, m_r\}$. This guides the code design and furthermore, the potential, of a linear binary JNCC satisfying some conditions, to achieve full diversity, can be verified without performing Monte Carlo simulations. 

We first prove that the diversity order is a function of only the network coding rate $R_n$ (Sec. \ref{sec: Diversity as a function of the network coding rate}). 
We then determine in Sec. \ref{sec: Space diversity by cooperation} the relation between the diversity order $d$ and the set $\{n_{u_r}, u_r=1, \ldots, m_r\}$, for \textit{any} linear binary JNCC expressed as in Eqs. (\ref{eq: general parity-check matrix form}) and (\ref{eq: H_GLNC}). The set $\{n_{u_r}, u_r=1, \ldots, m_r\}$ actually determines the maximal spatial diversity that can be achieved by cooperation, leading to an upper bound on the diversity order. In Sec. \ref{sec: Coding Matrix Rank Criterion}, we propose a lower bound on the diversity order in the case that $n_{u_r} = n =2$, which depends on all transmission sets $\{\T({u_r}), {u_r}=1, \ldots, m_r\}$. In Sec. \ref{sec: Diversity order with interuser failures}, we discuss how the diversity order is affected by interuser failures. Finally, in Sec. \ref{sec: diversity for layered}, we briefly comment on the diversity order in a layered construction, such as the OSI model. 

\subsection{Diversity as a function of the network coding rate}
\label{sec: Diversity as a function of the network coding rate}

We denote the maximum achievable diversity order by $d_\mmax$. We will determine $d_{\textrm{max}}$ in this section and show that it only depends on the network coding rate $R_n=\frac{m_s}{m_s+m_r}$. 

\begin{proposition}
\label{prop: div vs Rn}
Under ML decoding, the maximum diversity order $d_\mmax$ that can be achieved by any linear JNCC is
\begin{equation}
\label{eq:singleton}
d_\mmax = \left\{	
\begin{array}{lr}
 \lceil \frac{1+m_r}{2} \rceil & \textrm{, if } m_r \leq 2 m_s \\
1+m_r - m_s & \textrm{, if }  m_r > 2 m_s
\end{array}
\right. .
\end{equation}
\end{proposition}
\begin{proof}
See App. \ref{sec: Proof of Prop. div vs Rn}. 
\end{proof}

Note that the maximal diversity order does not depend on $L$. It can actually be reformulated in the following way:
\begin{equation}
\label{eq:singleton2}
d_\mmax = \left\{	
\begin{array}{lr}
 \lceil \frac{1+(1-R_n)(m_r+m_s)}{2} \rceil & \textrm{, if } m_r \leq 2 m_s \\
1+m_r - (m_s+m_r)R_n &  \textrm{, if } m_r > 2 m_s
\end{array}
\right. ,
\end{equation}
which for $m_r=m_s=m$ reduces to the maximum diversity order for a standard BF channel\footnote{A standard BF channel is a channel with $B$ blocks of length $L$, where each block is affected by an independent fading gain. The maximal achievable diversity order on this channel is given by $1+\lfloor B(1-R_c) \rfloor$, where $R_c$ is the coding rate \cite{knopp2000cbf,malkamaki1999epc, fab2007cmi}.} with $m$ blocks and coding rate $R_n$ \cite{knopp2000cbf,malkamaki1999epc, fab2007cmi}.

Hence, the maximum diversity order does not change when the point-to-point channel coding rate $R_{c,p}$ changes. This corresponds with our intuition as the parity bits of the point-to-point codes only provide redundancy within one block forming a point-to-point codeword, hence these parity bits cannot combat erasures which affect the complete point-to-point codeword. Another consequence is that the maximal diversity order of JNCC cannot be larger than in a layered approach, with the same network coding rate.

In the remainder of the paper, full diversity refers to the diversity order being equal to the maximal diversity order, $d=d_\mmax$, from (\ref{eq:singleton}).

\subsection{Space diversity by cooperation}
\label{sec: Space diversity by cooperation}

We denote the word error rate for each source ${u_s}$ by $P_{\ew,{u_s}}$, which is the fraction of packets where at least $1$ of the $K$ information bits from source ${u_s}$ is erroneously decoded at the destination. Associated to $P_{\ew,{u_s}}$, we define $d_{u_s}$, so that $P_{\ew,{u_s}} \propto \frac{1}{\gamma^{d_{u_s}}}$ for large $\gamma$. We have that $\max_u P_{\ew,{u_s}} \leq P_{\ew} \leq \sum_{u_s} P_{\ew,{u_s}}$. From Def. \ref{def: div block fading}, it follows that 
\begin{equation}
\label{eq: min du}
	d = \min_{u_s} d_{u_s}.
\end{equation}

Denote $t_{u_s}$, ${u_s} \in \{1,\ldots, m_s\}$, as the number of times that source ${u_s}$ is included in the transmission set of a relay: $t_{u_s} = \sum_{u_r \neq {u_s}} \ind({u_s} \in \T(u_r))$, where $\ind(.)$ is the indicator function, which equals one when its argument is true and zero otherwise. Some simple measures can be determined: $t_{min} = \min_{u_s} t_{u_s}$ and $t_{\textrm{av}} = \frac{\sum_{{u_r}=1}^{m_r} n_{u_r}}{m_s}$. We will show that $d_{u_s}$ depends on $t_{u_s}$ and thus, by Eq. (\ref{eq: min du}), $d$ depends on $t_{min}$. We denote $1 + t_{min}$ by $d_R$, which we call the space diversity order, as it is the minimal number of channels that convey a source message to the destination. 

\begin{proposition}
\label{prop: tmin}
For any linear JNCC, applied in our system model, the diversity order $d$ is upper bounded as
\[d \leq d_R = 1 + t_{min}.\]
\end{proposition}
\begin{proof}
We use the diversity equivalence between a BF channel and block BEC \cite{bou2011dac, bou2012tde}. Assume that the channel between source ${u_s}$ and the destination is erased. Source ${u_s}$ is included in at most $t_{u_s}$ transmission sets. Assume that all $t_{u_s}$ channels between the relays, that include source ${u_s}$ in their transmission set, and the destination are also erased. Then the destination does not receive any information on source ${u_s}$ so that it can never retrieve its message. The probability of occurrence of this event is $\epsilon^{1+t_{u_s}}$, so that $P_{\ew,{u_s}} \geq \epsilon^{1+t_{u_s}}$, hence $d_{u_s} \leq 1+t_{u_s}$. Using Eq. (\ref{eq: min du}), we obtain Prop. \ref{prop: tmin}. 
\end{proof}
Note that the proof of Prop. \ref{prop: tmin} is based on the assumption that relay ${u_r}$ only considers packets transmitted in the source phase for inclusion in $\Set({u_r})$. In the case that relay ${u_r}$ computes its relay packet also based on packets transmitted by other relays during the relay phase, the diversity order becomes more difficult to analyse.

In Cor. \ref{corr: new bound}, we propose the conditions on $t_{\textrm{min}}$ so that the space diversity order $d_R$ is not smaller than the maximum achievable diversity order.

\begin{corollary}
\label{corr: new bound}
	For any linear JNCC, applied in our system model, full diversity can be achieved only if $t_{\textrm{min}} \geq q$, where 
	\[q = \left\{ 
	\begin{array}{lr}
		\lf \frac{m_r}{2} \rf  & \textrm{, if } m_r \leq 2 m_s \\
		m_r-m_s & \textrm{, if } m_r > 2 m_s
	\end{array}
	\right. \]
\end{corollary}
\begin{proof}
The proof follows directly from Props. \ref{prop: div vs Rn} and \ref{prop: tmin}.
\end{proof}

Given a GLNC, and thus a choice of $\T({u_r})$, one can verify through Cor. \ref{corr: new bound} whether full diversity can be achieved. However, to get more insight for the code design, we consider the simplest case of a network code where the cardinality of the transmission set is constant ($n_{u_r}=n$).

\begin{corollary}
\label{corollary 2: n}
	For any linear JNCC, applied in our system model, with constant $n_{u_r}=n$, full diversity can be achieved only if 
	\begin{equation}	
		\left\{
		\begin{array}{lr}
			n \geq \lf \frac{m}{2} \rf & \textrm{, if } m_r=m_s=m \\
			n \geq \lc \frac{m_s}{2} \rc & \textrm{, if } 2 m_s \geq m_r > m_s \\
			n \geq m_s - \lf \frac{m_s^2}{m_r} \rf & \textrm{, if }  m_r > 2 m_s 
		\end{array}
		\right.
	\end{equation}
\end{corollary}
\begin{proof}
It always holds that $t_{\textrm{min}} \leq \lf t_{\textrm{av}} \rf$ and if $n_{u_r}=n$, then $t_{\textrm{av}} = \frac{m_r n}{m_s}$. 
From Cor. \ref{corr: new bound}, full diversity can be achieved only if $\lf \frac{m_r n}{m_s} \rf \geq q$. Because $\frac{m_r n}{m_s} \geq \lf \frac{m_r n}{m_s} \rf$, we have the necessary condition that $n \geq q \frac{m_s}{m_r}$. As $n$ is an integer, this bound can be tightened, yielding $n \geq \lc \frac{m_s}{m_r} q \rc$. Filling in $q$ from Cor. \ref{corr: new bound} yields Cor. \ref{corollary 2: n}.
\end{proof}

Table \ref{table: mr ms} illustrates Cor. \ref{corollary 2: n}, showing the set of networks in which a certain parameter $n$ is diversity-optimal, which means that the choice of $n$ does not prevent the code to achieve full diversity. In Sec. \ref{sec: practical JNCC for n_{u_r}=2}, we propose a JNCC for $n=2$, where taking $n=2$ is diversity-optimal in all networks corresponding to bold elements in Table \ref{table: mr ms}.

\begin{table}[!ht]
\centering
\begin{tabular}{|l|ccccccc|}
\hline
$m_r \backslash m_s$ & $1$ & $2$ & $3$ & $4$ & $5$ & $6$ & $7$ \\
\hline
$1$	      & $0$ &   &   &   &   &   &    \\									
$2$	      & $1$ &	$1$ &   &   &   &   &    \\								
$3$		  & $1$ & $1$ &	$1$ &   &   &   &    \\							
$4$		  & $1$ &	$1$ &	\textcolor{red}{$\mathbf{2}$} &	\textcolor{red}{$\mathbf{2}$} &   &   &    \\						
$5$		  & $1$ &	\textcolor{red}{$\mathbf{2}$} &	\textcolor{red}{$\mathbf{2}$} &	\textcolor{red}{$\mathbf{2}$} &	\textcolor{red}{$\mathbf{2}$} &   &    \\					
$6$		  & $1$ &	\textcolor{red}{$\mathbf{2}$} &	\textcolor{red}{$\mathbf{2}$} &	\textcolor{red}{$\mathbf{2}$} & $3$ &	$3$ &      \\
$7$		  &	$1$ &	\textcolor{red}{$\mathbf{2}$} &	\textcolor{red}{$\mathbf{2}$} &	\textcolor{red}{$\mathbf{2}$} &	$3$ &	$3$ &	$3$    \\
$8$		  &	$1$ &	\textcolor{red}{$\mathbf{2}$} &	\textcolor{red}{$\mathbf{2}$} &	\textcolor{red}{$\mathbf{2}$} &	$3$ &	$3$ &	$4$  \\	
\hline
\end{tabular}
\caption{Minimal value $n$ for a JNCC with constant $n_{u_r}=n$ to maintain its capability to achieve full diversity.}
\label{table: mr ms} 
\end{table}

\subsection{A lower bound based on $\{\T({u_r})\}$ for $n_{u_r}=2$}
\label{sec: Coding Matrix Rank Criterion}

A certain relay does not help one source only, but a combination of sources, expressed by the transmission set $\T(u_r)$ for each relay $u_r$. In this section we provide a lower bound on the diversity order, based on the choice of $\{\T({u_r}), u_r=1, \ldots, m_r\}$. If this lower bound and the upper bound in the previous section are tight, the exact diversity order of JNCCs can so be determined, as will be illustrated in Sec. \ref{sec: practical JNCC for n_{u_r}=2}.

Based on $\T({u_r}), m_s$ and $m_r$, we construct the $(m_s+m_r) \times m_s$ coding matrix $M$, where
\begin{equation}
\left\{
	\begin{array}{lr}
	M_{u_s,u_s}=1 & \textrm{for } u_s=1, \ldots, m_s \\
	M_{u_r+m_s,{u_s}}=1 & \textrm{if }{u_s}\in \T(u_r), \forall~u_s,u_r \\
	M_{i,{u_s}}=0  & \textrm{otherwise}
	\end{array}
\right.
\end{equation}
The matrix $M$ expresses the presence of a source-codeword in each transmission, i.e., $M_{i,{u_s}}=1$ if $\mb{s}_{u_s}$ is considered in transmission $i$ ($i = 1, \ldots, m_s$ and $i = m_s+1, \ldots, m_s+m_r$ correspond to the source and relay transmission phases, respectively). Therefore, the upper part of $M$ is an identity matrix as each source $u_s$ transmits its own codeword $\mb{s}_{u_s}$ in the source phase. The matrix $M$ represents what is often called the ``coding header'' or ``the global coding coefficients'' in the network coding literature (see e.g. \cite{chou2003practical}).

Consider a block BEC channel where $e$ of the $m_r$ blocks have been erased. The indices of the fading gains corresponding to the erased blocks are collected in the set $\mathcal{E}=\{ \mathcal{E}_1, \ldots, \mathcal{E}_e\}, \mathcal{E}_i \in \{1, \ldots, m_r\}$). Based on $\mathcal{E}$, we construct $M_{\mathcal{E}}$ which corresponds to the subset of transmissions that are not erased, i.e., all rows $\mathcal{E}_i$ (if $\mathcal{E}_i \leq m_s$) and $m_s+\mathcal{E}_i$, for $i=1, \ldots, e$, in $M$ are dropped. We denote the rank of $M_{\mathcal{E}}$ as $r_{M_{\mathcal{E}}}$. The set $\mathcal{M}(e)$ collects all possible matrices $M_{\mathcal{E}}$ which can be constructed from $M$ if $|\mathcal{E}|=e$.

Consider an example for $m_s = m_r =3$. Assume that $\T(1)=\{2,3\}$, $\T(2)=\{1,3\}$ and $\T(3)=\{1,2\}$, so that 
\begin{equation}
\label{eq: coding matrix m=3}
 M = 
\left[ \begin{array}{x{0.6cm}x{0.6cm}x{0.6cm}}
      1 & 0 & 0 \tn
      0 & 1 & 0 \tn
      0 & 0 & 1 \tn
      0 & 1 & 1 \tn
      1 & 0 & 1 \tn
      1 & 1 & 0
     \end{array} \right]
\end{equation}
Next, assume that $\mathcal{E}=\{1\}$. Hence, the channel between user $1$ and the destination is erased, so that rows $1$ and $4$ from $M$ are dropped: 
\begin{equation}
 M_{\mathcal{E}} = 
\left[ \begin{array}{x{0.6cm}x{0.6cm}x{0.6cm}}
      0 & 1 & 0 \tn
      0 & 0 & 1 \tn
      1 & 0 & 1 \tn
      1 & 1 & 0
     \end{array} \right],
\end{equation}
and $r_{M_{\mathcal{E}}} = 3$. It can be verified that all matrices $M_{\mathcal{E}} \in \mathcal{M}(1)$ have rank $r_{M_{\mathcal{E}}} = 3$. However, there exist matrices $M_{\mathcal{E}} \in \mathcal{M}(2)$ having rank $r_{M_{\mathcal{E}}} < 3$.

We can now define a metric that depends on $\{\T({u_r})\}$.
\begin{definition}
\label{def: d_M}
	We define $d_M = e^*+1$, where $e^*$ is the maximal cardinality of $\mathcal{E}$ such that $r_{M_{\mathcal{E}}} = m_s$ for each $M_{\mathcal{E}} \in \mathcal{M}(e)$.
\end{definition}
A simple computer program can compute $d_M$, given $\T({u_r}), m_s$ and $m_r$.

\begin{lemma}
\label{lemma: d_M}
	In a JNCC following the form of Eq. (\ref{eq: general parity-check matrix form}) with $m_s=m_r$ and constant $n_{u_r} = n =2$, the metric $d_M$ is at most three.
\end{lemma}
\begin{proof}
	If $m_s=m_r$ and $n=2$, then the minimum column weight of $M$ is smaller than or equal to three. Erasing the three rows where $M_{i,{u_s}}=1$, for a certain ${u_s}$ corresponding to the minimum column weight, leads to $M_{\mathcal{E}}$ having at least one zero column, and thus $r_{M_{\mathcal{E}}} < m_s$. By Def. \ref{def: d_M}, $d_M < 4$.
\end{proof}

In the next proposition, we provide a lower bound on the diversity order under ML decoding or Belief Propagation (BP) decoding \cite{McEliece1998tda}. We denote 
\begin{equation}
\label{eq: matrix with L equations}
 \mathcal{H}_{u_s}^{u_r} =  \left[ \begin{array}{c}
             H_p \\
	     	H_{u_s}^{u_r}
            \end{array}\right] , ~~ 
    \mathcal{H}_{u_r} = \left[ \begin{array}{c}
             H_p \\
	     	H_{u_r}
            \end{array}\right],
\end{equation}
which are square matrices of dimension $L$.

\begin{proposition}
\label{prop: coding matrix criterion}
	Using ML decoding, the diversity order of a JNCC following the form of Eq. (\ref{eq: general parity-check matrix form}) with constant $n_{u_r} = n =2$, is lower bounded as 
	\[d \geq d_M, \]
if the matrices $\mathcal{H}_{u_s}^{u_r}$, $u_s \in \T(u_r), u_r \in \{1, \ldots, m_s\}$, have full rank.

	 Using BP-decoding, the diversity order of a JNCC following the form of Eq. (\ref{eq: general parity-check matrix form}) with constant $n_{u_r} = n =2$, is lower bounded as 
	\[d \geq d_M, \]
if, for each ${u_r}$, the set of $L$ equations
\begin{equation}
\label{eq: useful rows 5}
\mathcal{H}_{u_r} \mb{r}_{u_r} = \bigoplus_{{u_s} \in \T({u_r})} \mathcal{H}_{u_s}^{u_r} \mb{s}_{u_s},
\end{equation}
can be solved with BP in the case of only one unknown source-codeword vector.
\end{proposition}
\begin{proof}
	See App. \ref{app: coding matrix criterion}.
\end{proof}

We can simplify the condition for BP decoding, stated in Prop. \ref{prop: coding matrix criterion}, when we assume that the parity bits of point-to-point codes do not have a support in $H_\GLNC$, or said differently, when the $L-K$ right most columns of the matrices $H_{u_r}$ and $H_{u_s}^{u_r}$ are zeroes. In that case, one iteration in the backward substitution, mentioned in App. \ref{app: coding matrix criterion}, corresponds to solving the $K$ unknown information bits of $\mb{s}_u$ via the set of $K$ equations 
\begin{equation}
\label{eq: relay BP solvability}
H_{u}^{u_r} \mb{s}_{u} = \bigoplus_{\substack{u_s \in \mathcal{T}({u_r})\\ u_s \neq u}} H_{u_s}^{u_r} \mb{s}_{u_s} + H_{u_r} \mb{y}_{m_s + {u_r}}.
\end{equation}
In Sec. \ref{sec: practical JNCC for n_{u_r}=2}, we propose a JNCC where the parity bits of point-to-point codes do not have a support in $H_\GLNC$, so that we take (\ref{eq: relay BP solvability}) instead of (\ref{eq: useful rows 5}) as condition for BP decoding in the remainder of the paper.

\subsection{Diversity order with interuser failures}
\label{sec: Diversity order with interuser failures}

It is often easier to prove that a particular diversity order is achieved assuming perfect interuser channels (see for example in Sec. \ref{sec: practical JNCC for n_{u_r}=2}). Here, we discuss how this diversity order is affected by interuser failures.

\begin{lemma}
\label{lemma: interuser failure general}
	In the case of non-reciprocal interuser channels, any JNCC achieves the same diversity order with or without interuser channel failures. 
\end{lemma}
\begin{proof}
	See Appendix \ref{app: proof of lemma interuser}. 
\end{proof}

In the case of reciprocal interuser channels, the achieved diversity order with interuser failures depends on the transmission sets $\{\T(u_r), u_r=1, \ldots, m_r\}$. We propose an algorithm to construct $\{\T(u_r)\}$ in Sec. \ref{sec: practical JNCC for n_{u_r}=2} and we will then discuss the diversity order with reciprocal interuser channels.

\subsection{Diversity order in a layered construction}
\label{sec: diversity for layered}

In a layered construction, such as the standard OSI model, the destination first attempts to decode the point-to-point transmissions. If it can not successfully retrieve the transmitted point-to-point codeword for a particular node-to-destination channel, then it declares a block erasure, where a block refers to one point-to-point codeword. Denoting this block erasure probability by $\epsilon$, we have that $\epsilon \propto \frac{1}{\gamma}$ \cite{tse2005fwc}. If for example $e$ blocks of length $L$ are erased, then the decoding corresponds to solving a set of equations with $eL$ unknowns.

Standard linear network coding consists of taking \textit{linear combinations} of several source packets. In general, non-binary coefficients are used in the linear combinations. Hence, packets are treated symbol-wise, which is shown to be capacity achieving for the layered construction \cite{koetter2003algebraic}. Hence, in Eq. (\ref{eq: relay}), the matrices $H_{u_r}$ and $H_{u_s}^{u_r}$ are replaced by identity matrices, which are multiplied with a non-binary coefficient in general. A consequence of this symbol-wise treatment is that the effective block length of the network code reduces to $m_s+m_r$ and the set of equations is expressed by the coding matrix $M_{\mathcal{E}}$. At this block length, ML decoding (which is equivalent to Gaussian elimination at the network layer) has low complexity. Therefore, non-random linear network codes being maximum distance separable (MDS) achieve the diversity order $d_M$ (Def. \ref{def: d_M}). Also note that random linear network codes are MDS codes with high probability for a sufficiently large field size \cite{ho2006random}. 

\section{Practical JNCC for $n_{u_r}=2$}
\label{sec: practical JNCC for n_{u_r}=2}

In the literature, a detailed diversity analysis is most often lacking. Codes were proposed and corresponding numerical results suggested that a certain diversity order was achieved on a \textit{specific} network. It is sometimes not clear why this diversity order is achieved, and how it would vary if the network or some parameters change. In the previous section, we made a detailed diversity analysis of a JNCC following the form of Eq. (\ref{eq: general parity-check matrix form}). However, the utility of for example Prop. \ref{prop: coding matrix criterion} is limited to JNCCs following the form of Eq. (\ref{eq: general parity-check matrix form}) with a constant $n_{u_r}=2$, which suggests that it is very hard to rigorously prove diversity claims in general. However, the modest analysis made in Sec. \ref{sec: diversity analysis of JNCC} can be applied in some cases and we will show its utility through an example. 

We consider networks with $m_s=m_r=m\geq 4$ and a JNCC following the form of Eq. (\ref{eq: general parity-check matrix form}) with $n_{u_r}=n=2$ for $u_r=1, \ldots, m$. We will rigorously prove that a diversity order of three is achieved, using the Props. of Sec. \ref{sec: diversity analysis of JNCC}. From Table \ref{table: mr ms}, it can be seen that this JNCC is diversity-optimal for $m=4$ and $m=5$. In Sec. \ref{sec: numerical results}, we provide numerical results for $m=5$.

From Table \ref{table: mr ms}, it is clear that restricting $n$ to two is not diversity-optimal in larger networks. However, it also has some advantages. If $n=2$, then every relay just needs to decode $2$ users, and encoding is restricted to taking a linear transformation of only two source packets. Furthermore, taking $n=2$ does not impose infeasible constraints on the number of sources in the vicinity of a relay in the case that spatial neighbourhoods are taken into account. Next, the theoretical analysis is simpler in the case $n=2$. Finally, taking $n=2$ allows to reuse strong codes designed for the multiple access relay channel, e.g. in \cite{duyck2010aac, duyck2009afj}.

Besides the diversity order, we indicated in Sec. \ref{sec: introduction} that scalability is also very important. The JNCC proposed here is scalable to any large network without requiring a redesign of the code. This means that we provide an on the fly construction method. The latter is particularly important for self regulating networks. As a node adds itself to the network, it can seamlessly integrate to the network. Together with the new symbols sent by the new node, a new JNCC code is formed which still possesses all desirable properties. Finally, note that due to the large block length of JNCC, ML decoding is too complex and low-complexity techniques, such as BP decoding, must be used. 

Hence, two properties are claimed: scalability to large networks and a diversity order of three (which is full diversity in some cases) under BP decoding. The JNCC code is presented in two steps. First, we present the design of $\{\T({u_r})\}$ and thus the coding matrix $M$. In a second step (Eq. (\ref{eq: example network advanced})), we specify the matrices $H_{u_r}$ and $H_{u_s}^{u_r}$ and we will prove that the scalability and the diversity order of three are achieved. 

\subsection{First step: design of $\T({u_r})$}
\label{sec: On the fly construction method}

The transmission sets $\{\T({u_r})\}$ have a large impact on the diversity order. For example, in \cite{duyck2011tfd}, a random construction was studied (each relay chooses $n=2$ sources at random) and it was shown that $\mathbb{E}[t_{u_s}]=2$, but $\textrm{Var}[t_{u_s}]=2$ as well, so that most probably $t_{\min}<2$ and $d_R < 3$ (Prop. \ref{prop: tmin}). So we need a more intelligent construction. 

We present an algorithm to determine $\{\T({u_r})\}$, given $m_s$ and $m_r$, and we subsequently determine the corresponding metrics $t_{\min}$ and $d_M$. We define the function $f_{m_s}(x) = \left( (x-1) \mod m_s \right) +1$ which adapts the modulo operation to the range $1\leq f_{m_s}(x) \leq m_s$.

\begin{algorithm}
\caption{Choose transmission set $\T({u_r})$}
\label{alg:1}
\begin{algorithmic}
\STATE 
\FOR{each relay ${u_r}=1\to m_r$}
	\STATE Set $u_1 =  f_{m_s}({u_r}+1)$
	\STATE Set $u_2 =  f_{m_s}({u_r}+2)$
	\STATE Set $\mathcal{S}({u_r})=\{u_1, u_2\}$
	\IF {$\{u_1, u_2  \} \subset \R({u_r})$}
	  \STATE $\T({u_r}) = \{u_1, u_2 \}$
	\ELSE
	  \STATE $\T({u_r}) = \{\}$: relay stays silent
	\ENDIF
\ENDFOR
\end{algorithmic}
\end{algorithm}

The transmission set $\T({u_r})$ is expressed via the bottom part of $M$. An example of such a matrix $M$ is given in Eq. (\ref{eq: example network XORing}) for $m_s=m_r=5$.
\begin{equation}
\label{eq: example network XORing}
 M = 
 \scriptsize
\left[ \begin{array}{x{0.4cm}x{0.4cm}x{0.4cm}x{0.4cm}x{0.4cm}}
      1 & 0 & 0 & 0 & 0 \tn
      0 & 1 & 0 & 0 & 0 \tn
      0 & 0 & 1 & 0 & 0 \tn
      0 & 0 & 0 & 1 & 0 \tn
      0 & 0 & 0 & 0 & 1 \tn
      0 & 1 & 1 & 0 & 0 \tn
      0 & 0 & 1 & 1 & 0 \tn
      0 & 0 & 0 & 1 & 1 \tn
      1 & 0 & 0 & 0 & 1 \tn
      1 & 1 & 0 & 0 & 0 \tn
  \end{array} \right]
\end{equation}

If a node is added as a source node, it adopts the largest source index, $m_s+1$, and relay-only nodes, with indices larger than or equal to $m_s+1$, increment their index by one. The function $f_{m_s}(x)$ is updated to the new $m_s$. Note that the algorithm corresponds to a deterministic cooperation strategy, which avoids extra signalling to the destination regarding the code design.

We first consider the case of perfect interuser channels and prove that Alg. \ref{alg:1} yields $d=3$ (Cor. \ref{cor: full diversity for n=2}). We then consider interuser failures and prove that the diversity order is not affected (Lemma \ref{lemma: one failure}).

\begin{corollary}
\label{cor: full diversity for n=2}
	Having perfect links from sources to relays, the diversity order of a JNCC, with $m_s=m_r$ and with transmission set constructed via Algorithm \ref{alg:1}, achieves a diversity order $d =3$ using BP-decoding, if, for each ${u_r}$, Eqs. (\ref{eq: relay BP solvability})
can be solved with BP in the case of only one unknown source-codeword vector.
\end{corollary}
\begin{proof}
	Because the links between sources and relays are perfect, the relays will never stay silent. In the case that $m_r = m_s$ and $n_{u_r} = 2$, we have that $t_{\min}=t_{\textrm{av}}=2$ and so $d_R=3$.
	
	Next, we show that $d_M=3$ (and thus, according to Lemma \ref{lemma: d_M}, $d_M$ is maximized if $n=2$). Consider $|\mathcal{E}|=2$. Without loss of generality, consider that $\mathcal{E}=\{1, 2\}$. Consider the set of equations $M_{\mathcal{E}} \mb{z}=\mb{c}$. Variables $z_3, \ldots, z_{m_s}$ can be recovered via the top $m_s-2$ rows of $M_{\mathcal{E}}$. The two relays $u_1$ and $u_2$ having source ${u_s}$ in their transmission set ($\T(u_1)$ and $\T(u_2)$, respectively) are 
\[u_1 = f_{m_s}({u_s}-1), u_2 = f_{m_s} ({u_s}-2).\]
Hence, source $1$ is included in $\T(m-1)$ and $\T(m)$, and source $2$ is included in $\T(m)$ and $\T(1)$. Hence, relay transmission $m-1$ can be used to retrieve source $1$ and relay transmission $m$ can be used to retrieve source $2$, as long as $m \geq 4$. Hence, $M_{\mathcal{E}}$ has full rank. The generalization to any set $\mathcal{E}$ satisfying $|\mathcal{E}|=2$, is straightforward. Therefore, we have that $d_M=3$.

	As $d_R=d_M=3$, the proof follows immediately from Props. \ref{prop: tmin} and \ref{prop: coding matrix criterion}. 
\end{proof}

Next, it can be proved that a JNCC applied in our system model has a diversity order of three, if it has a diversity order of three when all interuser channels are perfect. This is proved in general for non-reciprocal interuser channels in Lemma \ref{lemma: interuser failure general}, and here, we consider reciprocal interuser channels.

\begin{lemma}
\label{lemma: one failure}
	A JNCC, with transmission set constructed via Algorithm \ref{alg:1}, achieves the same diversity order with or without interuser channel failures when $m_s > 4$ or when $m_s=m_r=m \leq 4$. 
\end{lemma}
\begin{proof}
	See Appendix \ref{app: proof of lemma interuser2}. 
\end{proof}
For conciseness, we do not consider the other cases, $m_r > m_s \leq 4$.

\subsection{Second step: JNCC of LDPC-type}

In the first step, we specified $\{\T({u_r})\}$ and proved that $d_R=d_M=3$ if $m_r=m_s=m>3$. According to Cor. \ref{cor: full diversity for n=2}, a diversity order of three is achieved under BP decoding if, for each ${u_r}$, Eqs. (\ref{eq: relay BP solvability})
can be solved with BP in the case of only one unknown source-codeword vector. In the second step, we specify the sub matrices $H_{u_r}$, $H_{u_s}^{u_r}$, $\forall {u_r}, {u_s}$, to satisfy this condition, given that $\{\T({u_r})\}$ is constructed according to Alg. \ref{alg:1}.

A simple solution is to replace the $K$ left most columns in all $K \times L$ sub matrices $H_{u_r}$, $H_{u_s}^{u_r}$, $\forall {u_r}, {u_s}$, by identity matrices. In this case, the joint network channel coding essentially reduces to a layered solution: the source-codewords are decoded at the relays and simply added according to Eq. (\ref{eq: relay}). If the network code is used at the physical layer, it has to deal with noise and a more advanced code might be required. 

In the literature, a full-diversity close-to-outage performing JNCC for the Multiple Access Relay Channel (MARC) has been proposed \cite{duyck2009afj, duyck2010aac}, which is a code in the form of Eq. (\ref{eq: example JNCC}). 
These codes are such that the set of Eqs. 
\[H_{1}^1 \mb{s}_1  + H_{2}^1 \mb{s}_2 + H_{1} \mb{r}_1 = \mb{0} \]
can be decoded via BP if only one coding vector $\mb{s}_1$, $\mb{s}_2$ or $\mb{r}_1$ is erased and the other coding vectors are perfectly known. We denote this JNCC by MARC-JNCC. The matrix $H_{\textrm{GLNC, MARC}}$ of the MARC-JNCC is given by Eq. (A.7) in \cite{duyck2010aac}\footnote{The attentive reader will notice that the first two block rows in Eq. (A.7) in \cite{duyck2010aac} are not used here. These block rows are only necessary if a source is helped by one relay only and no point-to-point codes are available, which is not the case here.}:
\begin{equation}
\label{eq: proposal7 bis}
{\scriptsize H_{\textrm{GLNC, MARC}} = 
\stackrel{
\left.
\begin{array}{x{0.5cm}x{0.5cm}x{0.5cm}x{0.5cm}x{1.0cm}}
$\mb{1i_1}$ & $\mb{2i_1}$ & $\mb{1i_2}$ & $\mb{2i_2}$ & $\mb{r_1}$ 
\end{array}
\right.
}{
\left[ \begin{array}{x{0.5cm}x{0.5cm}x{0.5cm}x{0.5cm}x{1.0cm}}
\I & $R_1$ & \0 & \I & \multirow{2}{*}{$R_3$}  \tn
\0 & \I & \I & $R_2$ &   
\end{array}  \right]}}
\end{equation}
where $\mb{s}_j = [\mb{1i}_j ~ \mb{2i}_j ~ \mb{p}_j]$ is the codeword from source $j$, with $[\mb{1i}_j ~ \mb{2i}_j]$ and $\mb{p}_j$ denoting the information bits and the parity bits, respectively ($j=1,2$); $\mb{1i}_j$ and $\mb{2i}_j$ each contain $\frac{K}{2}$ information bits.
However, the parity bits $\mb{p}_j$ are not involved in $H_{\textrm{GLNC, MARC}}$. The matrices $R_i$, with $i=1,2,3$, are random matrices, chosen according to the required degree distributions of the LDPC code. To facilitate future notation, we denote 
\[
H_1 = {\scriptsize\left[ \begin{array}{x{0.45cm}x{0.45cm}x{0.3cm}}
\I & $R_1$ & \0 \tn
\0 & \I & \0  
\end{array}  \right]}, ~~ 
H_1^\prime = {\scriptsize\left[ \begin{array}{x{0.45cm}x{0.45cm}x{0.3cm}}
$R_1$ & \I & \0   \tn
\I & \0 & \0  
\end{array}  \right] },\]
\[
H_2 = {\scriptsize\left[ \begin{array}{x{0.45cm}x{0.45cm}x{0.3cm}}
\0 & \I & \0 \tn
\I & $R_2$ & \0  
\end{array}  \right]}, ~~
H_2^\prime = {\scriptsize\left[ \begin{array}{x{0.45cm}x{0.45cm}x{0.3cm}}
\I & \0 & \0 \tn
$R_2$ & \I & \0  
\end{array}  \right]}.\]
and $H_3 = R_3$, so that $H_{\textrm{GLNC}} = [\bar{H_1} ~ \bar{H_2} ~ H_3]$, where $\bar{H_i} = H_i \textrm{ or } H_i^\prime$ (it will become clear hereunder which one has to be chosen at each relay). In $\bar{H_1}$ and $\bar{H_2}$, the first two block columns each consist of $K/2$ columns (corresponding to information bits) and the last block column consists of $L-K$ columns (corresponding to parity bits from the point-to-point codes). The zero block columns indicate that parity bits from point-to-point codes have no support in these matrices. Now replace all sub matrices $H_{u_r}$, $H_{u_s}^{u_r}$ by these matrices, for each relay ${u_r}$, so that in each block column corresponding to information bits, we have a random matrix $R_i$; this is required to conform any preferred degree distribution of the LDPC code. For example, $H_{\textrm{GLNC}}$ can be given by
\begin{equation}
\label{eq: example network advanced}
\scriptsize
H_{\textrm{GLNC}} = 
\stackrel{
\left.
\begin{array}{x{0.37cm}x{0.37cm}x{0.37cm}x{0.37cm}x{0.37cm}x{0.37cm}x{0.37cm}x{0.37cm}x{0.37cm}x{0.37cm}}
$\mb{s}_1$ & $\mb{s}_2$ & $\mb{s}_3$ & $\mb{s}_4$ & $\mb{s}_5$ & $\mb{r}_1$ & $\mb{r}_2$ & $\mb{r}_3$ & $\mb{r}_4$ & $\mb{r}_5$
\end{array}
\right.
}{
\left[ \begin{array}{x{0.37cm}x{0.37cm}x{0.37cm}x{0.37cm}x{0.37cm}x{0.37cm}x{0.37cm}x{0.37cm}x{0.37cm}x{0.37cm}}
0 & $H_1$ & $H_2$ & 0 & 0 & $H_3$ & 0 & 0 & 0 & 0 \tn
0 & 0 & $H_1^\prime$ & $H_2^\prime$ & 0 & 0 & $H_3$ & 0 & 0 & 0 \tn
0 & 0 & 0 & $H_1$ & $H_2$ & 0 & 0 & $H_3$ & 0 & 0 \tn
$H_2^\prime$ & 0 & 0 & 0 & $H_1^\prime$ & 0 & 0 & 0 & $H_3$ & 0 \tn
$H_1$ & $H_2^\prime$ & 0 & 0 & 0 & 0 & 0 & 0 & 0 & $H_3$ 
\end{array}  \right]}
\normalsize
\end{equation} 
Each set of rows and each set of columns in $H$ will have at least one random matrix, so that any LDPC code degree distribution can be conformed. We denote this JNCC by the SMARC-JNCC, where S stands for scalable.

\begin{proposition}
\label{prop: second step}
	In a network following the system model proposed in Sec. \ref{sec: system model} and using BP, the SMARC-JNCC achieves a diversity order $d=3$. 
\end{proposition}
\begin{proof}
Consider the set of $K$ equations
\begin{equation}
\label{eq: prop code unknown source codeword}
H_3 \mb{r}_{u_r} = \bar{H_1} \mb{s}_{u_1} + \bar{H_2} \mb{s}_{u_2}, ~~\{u_1, u_2\} \in \T({u_r}).
\end{equation}
In \cite{duyck2010aac}, it is proved that this set of $K$ equations can be solved using the matrices proposed above. We provide another more simple proof here. Consider a block BEC. Because $\bar{H_1}$ and $\bar{H_2}$ are upper- or lower triangular, with ones on the diagonal, the unknown $K$ information bits can be retrieved using backward substition, hence it can be retrieved with BP as well. 

By Cor. \ref{cor: full diversity for n=2} and Lemma \ref{lemma: one failure}, the SMARC-JNCC achieves a diversity order $d=3$.
\end{proof}

Note that the information bits of a source need to be split in two parts: bits of the type $1i$ and $2i$. This allows the introduction of the matrices $R_{1}$ and $R_{2}$ in Eq. \ref{eq: proposal7 bis}, so that all information bits have a random matrix in their corresponding block column in the parity-check matrix. Now, the LDPC code can conform any degree distribution.

\section{Lower bound for the WER}
\label{sec: lower bound on the WER}

To assess the performance of the SMARC-JNCC we need to compare it with the outage probability limit (Sec. \ref{sec: Calculation of the outage probability}). We show that the outage probability limit is not always tight and we propose a tighter lower bound, which is presented in Sec. \ref{sec: Calculation of tight lower bound on WER}.

\subsection{Calculation of the outage probability}
\label{sec: Calculation of the outage probability}

The outage probability limit is the probability that the instantaneous mutual information between the sources and sinks of the network is less than the transmitted rate. The outage probability is an achievable (using a random codebook) lower bound of the average WER of coded systems in the limit of large block length \cite{biglieri1998fci, ozarow1994itc, fab2007cmi}.

For a multi-user environment, two types of mutual information are considered. First, it is verified whether the sum-rate, $R_c$ in this case, is smaller than the instantaneous mutual information between all the sources and the sink. Then, it is verified whether each individual source rate, $\frac{R_c}{m_s}$ in this case, is smaller than the instantaneous mutual information between the nodes, transmitting information for this source, and the destination. The outage probability for the MARC was determined in \cite{hausl2009jnc, duyck2010aac} using the method described above.

The outage probability is
\begin{equation*}
	P_{\textrm{out}} = P\big(\mathcal{E}_{\textrm{out}}),
\end{equation*}
where $\mathcal{E}_{\textrm{out}}$ is denoted as an outage event. Similarly as in \cite{hausl2009jnc, duyck2010aac}, an outage event is given by
\begin{align*}
	&  \mathcal{E}_{\textrm{out}} = \bigg\{ \left[ R_c \geq \frac{\sum_{u_s=1}^{m_s} I(S_{u_s};D) + \sum_{u_r=1}^{m_r} B_{u_r} I(R_{u_r};D)}{m_s+m_r}  \right] \\
		&  \cup_{u_s=1}^{m_s} \left[ \frac{R_c}{m_s} \geq \frac{I(S_{u_s};D) + \sum_{j | u_s \in \T(j)} B_{j} I(R_{j};D)}{m_s+m_r}  \right] \bigg\},
\end{align*}
where 
\begin{equation*}
B_j = \prod_{i \in \T(j)} \ind{\left[ I(S_i;R_j) > R_{c,p}  \right]}.
\end{equation*}
The terms $I(S_i;D)$, $I(R_i;D)$ and $I(S_i;R_j)$ are the instantaneous mutual informations of the corresponding point-to-point channels with input $x \in \{-1, 1\}$, received signal $y = \alpha_i x + w$ with $w \sim \mathcal{CN}(0,\frac{1}{\gamma})$, conditioned on the channel realization $\alpha_i$, which are determined by applying the formula for mutual information \cite{ungerboeck1982ccm, cover2006eit}:
\begin{equation*}
	I(X;Y|\alpha_i) = 1-\mathbb{E}_{Y|\{x=1,\alpha_i\}}\left\{\log_2\left(1+\exp\left[-4 y \alpha_i \gamma\right]\right)\right\},
	\label{eqn: mut info BPSK}
\end{equation*}
\noindent where $\mathbb{E}_{Y|\{x=1,\alpha_i\}}$ is the mathematical expectation over $Y$ given $x=1$ and $\alpha_i$. 

We now consider the outage probability of a layered construction, such as the standard OSI model, where the destination first decodes the point-to-point transmissions, declaring a block erasure if decoding is not successful. For the network code, we assume a maximum distance separable (MDS) code, which is outage-achieving over the (noiseless) block-erasure channel \cite{fabregas2006cit}. That is, any $m_s$ correctly received packets suffice for decoding. Accordingly, an outage event for the layered construction, denoted as $\mathcal{E}_{\textrm{out},l}$ is given by
\begin{align*}
	& \mathcal{E}_{\textrm{out},l} = \bigg\{ \left[ \sum_{u_s=1}^{m_s} E_{s, u_s} + \sum_{u_r=1}^{m_r} E_{r, u_r} > m_r \right] \\
		& \cup_{u_s=1}^{m_s} \left[ 1-E_{s, u_s} + \sum_{j | u_s \in \T(j)} (1-E_{r,j}) = 0  \right] \bigg\},
\end{align*}
where
\begin{equation*}
E_{s, u_s} = \ind{ \left[ I(S_{u_s};D) < R_{c,p}  \right] }
\end{equation*}
and 
\begin{equation*}
E_{r, u_r} = 1-B_{u_r} \ind{ \left[ I(R_j;D) > R_{c,p}  \right] }
\end{equation*}

\begin{figure}[!]
   \centering
   {\includegraphics[height = 0.5 \textwidth, angle = -90]{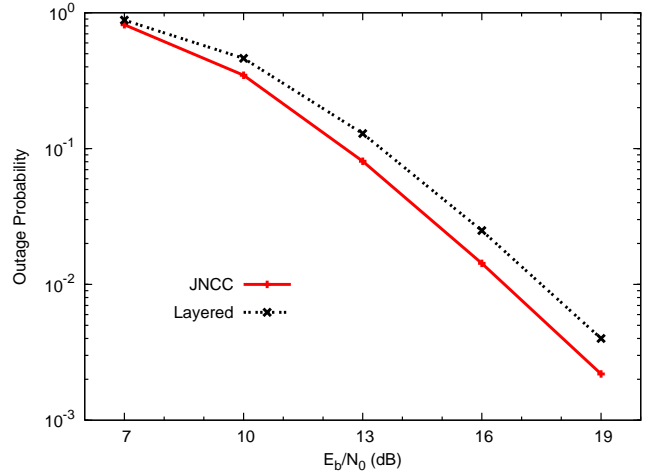}}
   \caption{The outage probabilities of JNCC and a layered construction are compared. The spectral efficiency is $R=3/7$ bpcu.}
	\label{fig: compare outage layered non-layered}
\end{figure}

The outage probability for JNCC and a layered construction are compared in Fig. \ref{fig: compare outage layered non-layered} for $m_s=m_r=5$, coding matrix\footnote{The coding matrix expresses the transmission sets for each relay, which is required to determine the outage probability.} $M$ given in Eq. (\ref{eq: example network XORing}) and $R_{c,p}=6/7$. The overall spectral efficiency is $R=3/7$ bpcu, so that $E_b/N_0 = \frac{7 \gamma}{3}$. 
The main conclusion is that the difference between both outage probabilities is only $1$dB. Hence, on a fundamental level, the achievable coding gain by JNCC with respect to a standard layered construction is small for the adopted system model.

\subsection{Calculation of a tighter lower bound on WER}
\label{sec: Calculation of tight lower bound on WER}

According to information theory, the outage probability is achievable, where the proof relies on using random codebooks. However, the nature of the JNCC protocol largely deviates from a random code. For example, the parity bits corresponding to the point-to-point codes are forced in a block diagonal structure in $H_c$ (see Eq. \ref{eq: general parity-check matrix form}), which is not taken into account in the outage probability limit. In fact, in Prop. \ref{prop: div vs Rn}, it was proved that the maximal diversity order does not depend on $R_c$ but on $R_n$, which is not taken into account in the outage probability limit. Therefore, we argue that the outage probability limit is in general not achievable by a JNCC, which is illustrated by means of an example. 

\begin{figure}[!]
   \centering
   {\includegraphics[height = 0.5 \textwidth, angle = -90]{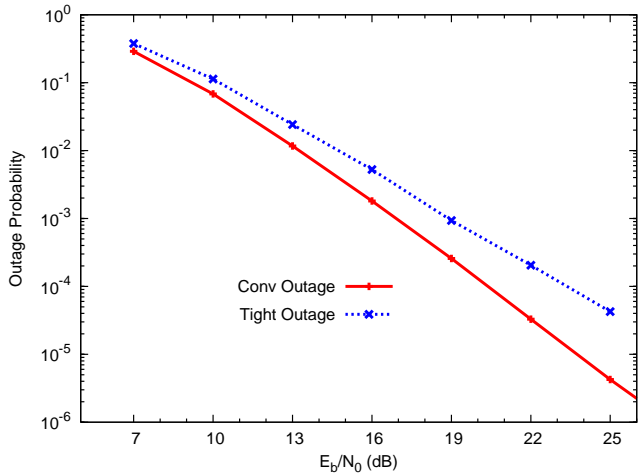}}
   \caption{The conventional and tighter outage probability of JNCC are compared.}
	\label{fig: outage m=3}
\end{figure}

Consider a network with $m_s=m_r=3$. The adopted point-to-point codes have coding rate $R_{c,p}=0.5$, so that $R_c=0.25$. We take $n_u=2$ and adopt the coding matrix $M$, given in Eq. (\ref{eq: coding matrix m=3}). Because of the small coding rate $R_c$, the outage probability achieves a diversity order of three (Fig. \ref{fig: outage m=3}). However, it follows from Prop. \ref{prop: div vs Rn} that $d_\mmax=2$. We therefore propose a new lower bound, which takes into account the point-to-point codes.

\begin{figure}[!]
   \centering
   {\includegraphics[width = 0.45 \textwidth]{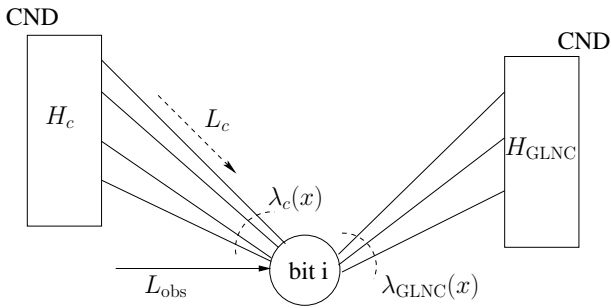}}
   \caption{The depicted part of the factor graph (using a Tanner notation) illustrates that a bit node (bit i on the figure) is essentially connected to two sets of check nodes, corresponding with $H_c$ and $H_\GLNC$, respectively. A set of check nodes is denoted as CND for check node decoder. The LLR-value coming from the CND corresponding with $H_c$ is denoted as $L_c$. The LLR-value corresponding with the channel observation is denoted as $L_{\textrm{obs}}$. }
	\label{fig: tight outage tanner 1}
\end{figure}

A bit node is essentially protected by two codes: a point-to-point code ($H_c$) and a network code ($H_\GLNC$), which is illustrated on the factor graph \cite{kschischang2001fga} representation (a Tanner notation \cite{tanner1981ara} is adopted)\footnote{For a specific instance, the parity-check matrix can be graphically represented by a bipartite graph, denoted as a Tanner graph. The graphical Tanner graph representation is equivalent to the factor graph, which can be used for decoding.} of the decoder (Fig. \ref{fig: tight outage tanner 1}). 
Usually, both codes are characterized by separate degree distributions, denoted as $(\lambda_c(x), \rho_c(x))$ and $(\lambda_\GLNC(x), \rho_\GLNC(x))$ for $H_c$ and $H_\GLNC$, respectively.

\begin{figure}[!]
   \centering
   {\includegraphics[width = 0.45 \textwidth]{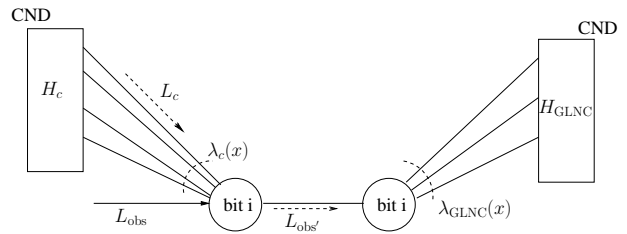}}
   \caption{The bit node in Fig. \ref{fig: tight outage tanner 1} can be duplicated with a single edge between both nodes as shown in this figure. The LLR $L_{\textrm{obs}^\prime}$ is the sum of all incoming LLR-values from the left, and contains the soft information which is passed to the network code decoder in a concatenated coding scheme. }
	\label{fig: tight outage tanner 2}
\end{figure}

The new lower bound assumes a concatenated decoding scheme. At the destination, first the point-to-point codes are decoded and then \textit{soft} information is passed to the network decoder. This is illustrated in Fig. \ref{fig: tight outage tanner 2}, where the soft information is denoted by the log-likelihood ratio (LLR) $L_{\textrm{obs}^\prime}$. Note that the bit node of bit i is duplicated to be able to clearly indicate $L_{\textrm{obs}^\prime}$. Applying the sum-product algorithm (SPA) on this factor graph or the original factor graph (without node duplication) is equivalent (see \cite{wymeersch2007ird} for a background on factor graphs and the SPA).
The LLR $L_{\textrm{obs}^\prime}$ can be viewed as a \textit{new} channel observation as it remains fixed during the iterative decoding of the network code ($H_\GLNC$). The maximum rate that can be achieved by the network code is given by 
\[\frac{1}{m_s+m_r} \left( \sum_{u_s=1}^{m_s} I(S_{u_s};L_{\textrm{obs}^\prime}) + \sum_{u_r=1}^{m_r} B_{u_r} I(R_{u_r};L_{\textrm{obs}^\prime}) \right).\]
The terms $I(S_i;L_{\textrm{obs}^\prime})$ and $I(R_i;L_{\textrm{obs}^\prime})$ are the mutual informations between the channel input $x \in \{-1, 1\}$ and the random variable $L_{\textrm{obs}^\prime}$, conditioned on the channel realization $\alpha_i$, determined by applying the formula for mutual information \cite{ungerboeck1982ccm, cover2006eit}, i.e., $I(X;L_{\textrm{obs}^\prime}|\alpha_i)$ is
\begin{equation*}
1-\mathbb{E}_{L_{\textrm{obs}^\prime}|\{x=1,\alpha_i\}}
	\left\{
	\log_2\left(1+ \frac{p_{L_{\textrm{obs}^\prime}}(l|x=-1,\alpha_i)}{p_{L_{\textrm{obs}^\prime}}(l|x=1,\alpha_i)} \right)\right\},
	\label{eqn: mut info BPSK L-value}
\end{equation*}
The density of the random variable $L_{\textrm{obs}^\prime}$ can be obtained by means of density evolution \cite{richardson2008mct}, given the degree distributions of the point-to-point code, or by means of Monte Carlo simulations, given the actual factor graph of the point-to-point code. Both approaches yield to the same results in our simulations. 

Similarly to the conventional case, an outage event, denoted as $\mathcal{E}_{\textrm{out},2}$ is given by
\begin{align*}
	& \textstyle \mathcal{E}_{\textrm{out},2} = \bigg\{ \left[ R_n \geq \frac{\sum_{u_s=1}^{m_s} I(S_{u_s};L_{\textrm{obs}^\prime}) + \sum_{u_r=1}^{m_r} B_{u_r} I(R_{u_r};L_{\textrm{obs}^\prime})}{m_s+m_r} \right] \\
		& \textstyle \cup_{u_s=1}^{m_s} \left[ \frac{R_n}{m_s} \geq \frac{I(S_{u_s};L_{\textrm{obs}^\prime}) + \sum_{j : u_s \in \T(j)} B_{j} I(R_{j};L_{\textrm{obs}^\prime})}{m_s+m_r}  \right] \bigg\}.
\end{align*}
Note that the network coding rate is used instead of the overall rate $R_c$, which corresponds to Prop. \ref{prop: div vs Rn}.

The tight lower bound presented here is a valid lower bound if the point-to-point codes are first decoded, followed by the network code, without iterating back to the point-to-point codes. 

Let us now go back to the small network example with $m_s=m_r=3$, considered in the beginning of this section. Fig. \ref{fig: outage m=3} compares the conventional outage probability (Sec. \ref{sec: Calculation of the outage probability}) with the tighter lower bound proposed here. As mentioned before, the conventional outage probability has a larger diversity order than what is achievable, while the tighter lower bound only achieves a diversity order of two. 
We are seeing a $3$dB difference at an outage probability of $10^{-4}$. To assess the performance of the network code only, given a certain point-to-point code, the WER of the SMARC-JNCC should be compared with the tight lower bound presented here. In the subsequent sections, we always include both lower bounds. 

\section{Numerical results}
\label{sec: numerical results}

\begin{figure}[!hpt]
\centering
	\includegraphics[width = 0.15\textwidth]{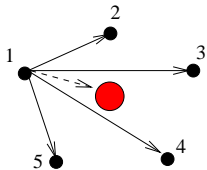}
\caption{The network example that will be used in this document is illustrated. The solid lines represent interuser channels, the dashed line is the channel to the destination. Only the channels from the perspective of user $1$ are shown for clarity, but all other users see equivalent channels.}
\label{fig: network example}
\end{figure}

In this section, we provide numerical results for the SMARC-JNCC. We will clarify the proposed techniques on an illustrating network example, where $m_s=m_r=5$ (Fig. \ref{fig: network example}). We use the same network example as in \cite{Bao2007gan, duyck2011tfd} so that a comparison is possible. 
For simplicity, we assume non-reciprocal interuser channel in the simulation results. Note that in the case that $m_s > 4$ and Alg. \ref{alg:1} is used to construct $\{\T(u_r), u_r=1, \ldots, m_r\}$, reciprocity is irrelevant for our proposed code, as it applies that $i \notin \T(j)$ if $j \in \T(i)$. 

We compare the error rate performance of the SMARC-JNCC with the outage probability limit and the tighter lower bound, which are presented in Sec. \ref{sec: lower bound on the WER}, and with standard network coding techniques (using identity matrices in $H_{\GLNC}$) and a layered network construction (also using identity matrices in $H_{\GLNC}$, and where, at the destination, the network code is only decoded after decoding all point-to-point codewords separately and taking a hard decision).

The point-to-point code used in the simulations is an irregular LDPC code \cite{richardson2008mct} characterized by the standard polynomials $\lambda(x)$ and $\rho(x)$ \cite{richardson2008mct}:
\begin{equation*}
	\lambda(x) = \sum_{i = 2}^{d_{b}}{\lambda_i x^{i-1}} ,~~~~	\rho(x) = \sum_{i=2}^{d_{c}}{\rho_i x^{i-1}}.
\end{equation*}
where $\lambda(x)$ and $\rho(x)$ are the left and right degree distributions from an edge perspective. The coefficients $\lambda_i$ and $\rho_i$ are the fraction of edges connected to a bit node and check node, respectively, of degree $i$. The adopted point-to-point code is fetched from \cite{hamdani2007cos}, has coding rate $R_{c,p}=6/7$ and conforms the following degree distributions:
\begin{align*}
& \lambda_2=0.173,~ \lambda_3=0.223,~ \lambda_4=0.095,~ \lambda_5=0.51 \\
& \rho_{24}=0.96,~ \rho_{25}=0.04.
\end{align*}

\subsection{Perfect source-relay links}
\label{sec: Perfect source-relay links}

We start by assessing the performance of $H_{\GLNC}$, the bottom part of Eq. (\ref{eq: example network advanced}), which determines the diversity order. Therefore, we assume perfect links between sources and relays. Hence, the channel model is the same as described in Sec. \ref{sec: system model}, with the exception of the interuser channels, which are assumed to be perfect (no fading and no noise). The parameters used for the simulation are $K=L=900$, $m_s=m_r=5$ (so that $N = 10 K = 9000$), where $N$ is the block length of the overall codeword. The overall spectral efficiency is $R=0.5$ bpcu, so that $E_b/N_0 = 2 \gamma$. 

\begin{figure}[!]
   \centering
   {\includegraphics[height = 0.5 \textwidth, angle = -90]{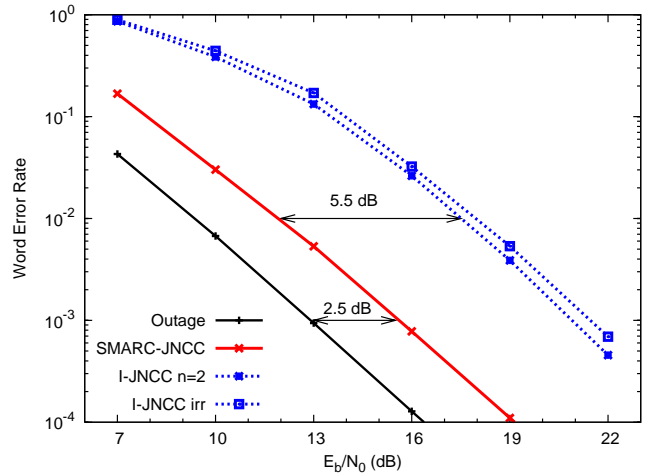}}
   \caption{The word error rate of the SMARC-JNCC is compared to that of the I-JNCC, assuming perfect source-relay channels.}
	\label{fig: perfect interuser channels}
\end{figure}

Fig. \ref{fig: perfect interuser channels} shows that a diversity order of $3$ is achieved for SMARC-JNCC, which corroborates Cor. \ref{cor: full diversity for n=2}. It performs at $2.5$dB from the outage probability (because no point-to-point codes are considered, only the conventional outage probability can be calculated), which may be improved by optimizing the degree distributions. We also show a JNCC, where all submatrices $H_{u_r}$, $H_{u_s}^{u_r}$, $\forall {u_r}, {u_s}$ are replaced by identity matrices, denoted as the I-JNCC. Finally, we show an I-JNCC with irregular $\{n_{u_r}\}$, with coding matrix $M$, given by
\begin{equation}
\label{eq: example network XORing I-JNCC random}
 M = 
 \scriptsize
\left[ \begin{array}{x{0.4cm}x{0.4cm}x{0.4cm}x{0.4cm}x{0.4cm}}
      1 & 0 & 0 & 0 & 0 \tn
      0 & 1 & 0 & 0 & 0 \tn
      0 & 0 & 1 & 0 & 0 \tn
      0 & 0 & 0 & 1 & 0 \tn
      0 & 0 & 0 & 0 & 1 \tn
      0 & 1 & 1 & 1 & 0 \tn
      0 & 1 & 1 & 0 & 1 \tn
      1 & 0 & 1 & 1 & 1 \tn
      1 & 1 & 1 & 1 & 1 \tn
      0 & 1 & 0 & 1 & 0 \tn
  \end{array} \right].
\end{equation}
It is clear that, even without optimizing the SMARC-JNCC, there is a benefit in terms of coding gain compared to the I-JNCC. 

\subsection{Rayleigh faded source-relay links}
\label{sec: rayleigh source-relay links}

Now, we assess the performance of the complete parity-check matrix $H$ of the SMARC-JNCC. We use the channel model as described in Sec. \ref{sec: system model}. Hence, all links have the same statistical model and the average SNR is the same as for all channels. The parameters used for the simulation are $K=606$, $R_{c,p}=6/7$, $L=707$, $m_s=m_r=5$ (so that $N = 10 L = 7070$). The overall spectral efficiency is $R=3/7$ bpcu, so that $E_b/N_0 = 7\gamma/3$. Because the simulation time would be very large if every point-to-point source-relay link has to be decoded separately, we made an approximation. The word error rate of the point-to-point code when transmitted on a channel with fading gain $\alpha$ is smaller than $10^{-4}$ when $\alpha^2 \gamma = 5.5$dB. Therefore, we assumed that a relay had correctly decoded the source-codeword if $\alpha^2 \gamma > 5.5$dB and not otherwise. We also add the performance of the SMARC-JNCC from Sec. \ref{sec: Perfect source-relay links}, corresponding to perfect source-relay links and $R=0.5$ bpcu, as a reference curve (note that the reference curve corresponds to a larger spectral efficiency - the coding rate $R_c$ is larger - than for the other curves, which slightly disadvantages the reference curve in terms of error performance). 

\begin{figure}[!]
   \centering
   {\includegraphics[height = 0.5 \textwidth, angle = -90]{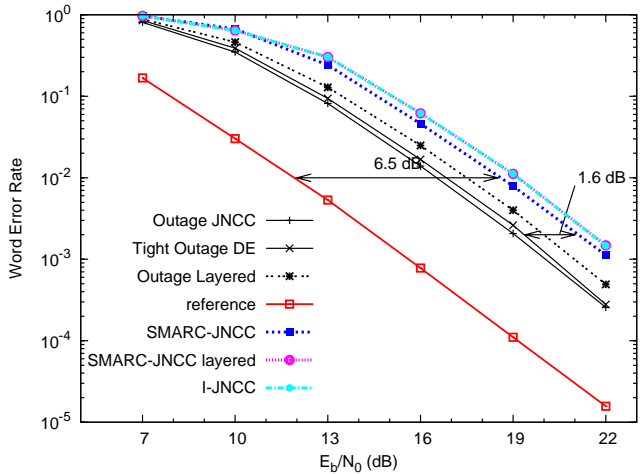}}
   \caption{The word error rate of the SMARC-JNCC is compared to that of the I-JNCC and a layered construction, assuming Rayleigh faded source-relay channels. The reference curve is the performance of the SMARC-JNCC assuming perfect source-relay channels (Sec. \ref{sec: Perfect source-relay links}).}
	\label{fig: rayleigh interuser channels}
\end{figure}

Fig. \ref{fig: rayleigh interuser channels} shows that a diversity order of $3$ is still achieved, which corroborates Prop. \ref{prop: second step}. In addition, two main conclusions can be made. First of all, the loss due to interuser failures is $6.5$dB, which is very large. Secondly, the benefit in terms of coding gain of the SMARC-JNCC compared to the I-JNCC is considerably decreased, compared to Sec. \ref{sec: Perfect source-relay links}, which corresponds to the small horizontal SNR-gap between the outage probabilities of a layered and joint construction. Also note that the tighter lower bound using density evolution, is close to the conventional lower bound in this case. Finally, the WER performance of a layered construction is shown, which coincides with that of the I-JNCC. 

\subsection{Gaussian source-relay links}

We test again the complete parity-check matrix $H$ of the SMARC-JNCC, now assuming that the source-relay links are Gaussian, having additive white Gaussian noise only, without fading; fading occurs on the source-destination and relay-destination links only. We assume that the average SNR is the same for all channels. The parameters used for the simulation are the same as in Sec. \ref{sec: rayleigh source-relay links}. 

\begin{figure}[!]
   \centering
   {\includegraphics[height = 0.5 \textwidth, angle = -90]{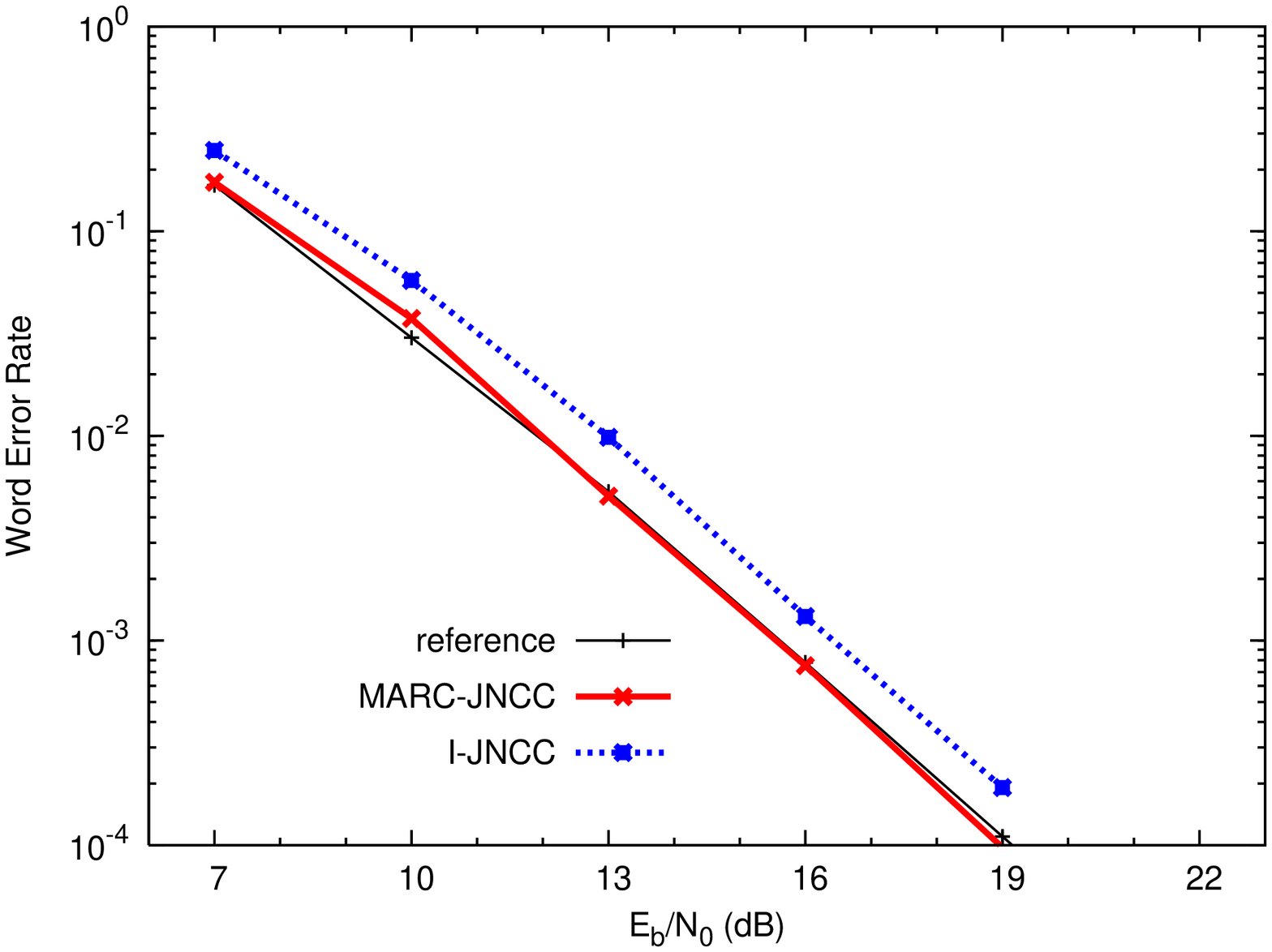}}
   \caption{The word error rate of the SMARC-JNCC is compared to that of the I-JNCC, assuming Gaussian source-relay channels. The reference curve is the performance of the SMARC-JNCC assuming perfect source-relay channels (Sec. \ref{sec: Perfect source-relay links}).}
	\label{fig: Gaussian interuser channels}
\end{figure}

Fig. \ref{fig: Gaussian interuser channels} shows that in the case of Gaussian interuser channels, the loss compared to perfect interuser channels is very small. Furthermore, the performance of the I-JNCC has improved a lot in comparison with Sec. \ref{sec: Perfect source-relay links} where $H_{\GLNC}$ only was used. The degree distributions causing the poor coding gain of the I-JNCC in Sec. \ref{sec: Perfect source-relay links}, are considerably improved by the point-to-point codes. 

\section{Conclusion}

We put forward a general form of joint network-channel codes (JNCCs) for a wireless communication network where sources also act as relay. The influence of important parameters of the JNCC on the diversity order is studied and an upper and lower bound on the diversity order are proposed. The lower bound is only valid for the case where the number of sources is equal to the number of relays, and where each relay only helps two sources. 

We then proposed a practical JNCC that is scalable to large networks. Using the diversity analysis, we managed to rigorously prove its achieved diversity order, which is optimal in a well identified set of wireless networks. We verified the performance of a regular LDPC code via numerical simulations, which suggest that as networks grow, it is difficult to perform significantly better than a standard layered construction.

\section*{Acknowledgement}

This work was supported by the European Commission in the framework of the FP7 Network of Excellence in Wireless COMmunications NEWCOM++ (contract n. 216715).

\appendix
\subsection{Proof of Prop. \ref{prop: div vs Rn}}
\label{sec: Proof of Prop. div vs Rn}

The maximal diversity order can be derived using the diversity equivalence between a block BEC and a BF channel \cite{bou2011dac, bou2012tde}. Assume a block BEC, so that a block $\mb{s}_{u_s}$ or $\mb{r}_{u_r}$ is completely erased or perfectly known. Consider the case that $e_1$ blocks of length $2L$ and $e_2$ blocks of length $L$ have been erased, where $e=e_1+e_2$ is the total number of erasures, $e_1 \leq m_s$ and $e_2 \leq m_r-m_s$. Hence, the number of unknown bits is equal to $e_1 2 L + e_2 L$. Considering the structure of $H$ from (\ref{eq: general parity-check matrix form}) containing the block-diagonal matrix $H_c$, it follows that the $e_1 2L+e_2 L$ erased bits appear in only $(2e_1+e_2)(L-K) + m_r K$ of the available $(m_s +m_r)L-m_s K$ parity equations, i.e., $(2e_1+e_2)(L-K)$ equations involving $H_c$ and all $m_rK$ equations involving $H_\GLNC$. Hence, the unknown bits can be retrieved only if there are sufficient linearly independent useful equations. This yields the necessary condition:
\begin{equation}
\label{eq: erasure constraint}
m_r \geq 2e_1+e_2.
\end{equation}
Denoting by $e = e_1+e_2$ the total number of erased blocks, the largest value $e_{\textrm{max}}$ of $e$ for which $e_1$ and $e_2$ satisfy (\ref{eq: erasure constraint}) for all $e_1 \leq m_s$ and $e_2 \leq m_r-m_s$ is given by
\begin{equation}
\label{eq: erasure constraint 2}
e_{\textrm{max}} = \left\{
\begin{array}{lr}
\lf \frac{m_r}{2} \rf & m_r \leq 2 m_s \\
m_r-m_s  & m_r > 2 m_s
\end{array}
\right.
\end{equation}
Hence, $d_{\textrm{max}}=e_{\textrm{max}}+1$, yielding Prop. \ref{prop: div vs Rn}.

%
%

\subsection{Proof of Prop. \ref{prop: coding matrix criterion}}
\label{app: coding matrix criterion}

Before we present the actual proof, we first propose two lemmas.

\begin{lemma}
\label{lemma: fullrank}
 Any binary $a \times b$ matrix $S$, $a \geq b$, where all rows have weight $2$ cannot have full rank $b$.
\end{lemma}
\begin{proof}
 If a matrix has full rank, there is no vector $\mb z \not =\mb 0$ such that $S \mb z = \mb 0$.
However, if $S$ has row weight $2$, then $S \mb 1 = \mb 0$, where $\mb 1$ corresponds to a column vector with each entry equal to $1$. 
\end{proof}

Consider now a column vector of $b$ unknown variables $\mb{z}$ and a set of constraints on these variables, which are stacked in $S$ so that $S \mb{z} = \mb{c}$, where $\mb{c}$ is a column vector of known constants. In general, solving $S$ for $\mb{z}$ corresponds to performing Gaussian elimination of $S$. However, under some conditions, this simplifies to backward substitution.

\begin{lemma}
 \label{lemma: BP}
If a binary $a \times b$ matrix $S$, $a \geq b$, has full rank $b$ and maximal row weight of $2$, Gaussian elimination simplifies to backward substitution.
\end{lemma}
\begin{proof}
Without loss of generality, we eliminate all redundant (linearly dependent) rows in $S$ to obtain a square matrix of size $b$. By Lemma \ref{lemma: fullrank}, there must be at least one row in $S$ with unit weight to have full rank. Starting from this known variable, we can solve for a further variable in $\mb{z}$ at each step as the row weight is smaller than or equal to $2$.

Assume that this backward substitution procedure cannot be continued until all variables are known. That is, after successive decoding, there are $k$ rows consisting of a combination of $\mb z_{i_k} + \mb z_{j_k}$ where neither $\mb z_{i_k}$ nor $\mb z_{j_k}$ are known. 
We split the matrix $S$ into two parts: $S_{\textrm{unknown}} \in \{0,1\}^{k \times m_s}$ and $S_{\textrm{known}} \in \{0,1\}^{m_s-k \times m_s}$. The former comprises the rows involving only unknown variables (note that the weight of each row of $S_{\textrm{unknown}}$ is $2$). The latter consists of the rows involving only known variables. 
If the number of unknown variables is equal to $k$, then the rank of $S_{\textrm{unknown}}$ must be equal to $k$ which is impossible by Lemma \ref{lemma: fullrank}. So, the matrix $S$ was not full rank which contradicts our assumption.
If the number of unknown variables is smaller than $k$, then there were redundant (linearly dependent) rows in $S_{\textrm{known}}$ which contradicts the assumptions again. We conclude that the procedure only fails if $S$ does not have full rank.
\end{proof}

To prove Prop. \ref{prop: coding matrix criterion}, we use the diversity equivalence between a block BEC and the BF channel. In a block BEC, the channel equation (\ref{eq: channel equation}) simplifies to 
\begin{equation}
\left\{
	\begin{array}{lr}
	\mb{y}_{u_s} = \epsilon_{u_s} \mb{s}_{u_s}^\prime, &  u_s=1, \ldots, m_s \\
	\mb{y}_{m_s+u_r} = \epsilon_{u_r} \mb{r}_{u_r}^\prime, & u_r =1, \ldots, m_r
	\end{array}
	\right.
\end{equation}
where $\epsilon_i=0$ when the channel is erased and $\epsilon_i=1$ otherwise. Hence, $\epsilon_i=0$ if $i \in \mathcal{E}$ and $\epsilon_i=1$ if $i \in \bar{\mathcal{E}}$, where $\bar{\mathcal{E}}$ is the complement of $\mathcal{E}$. 

Source-codewords $\mb{s}_i$ can be retrieved from the transmissions in the source phase if $\epsilon_i=0$. Decoding the other source-codewords at the destination is performed through the parity-check matrix $H$ (Eq. (\ref{eq: general parity-check matrix form})). 
We split $H$ in two parts:
\begin{equation}
	H = \left[
	\begin{array}{cc}
	H_{\textrm{left}} & H_{\textrm{right}} 
	\end{array}
	\right],
\end{equation}
where $H_{\textrm{left}}$ and $H_{\textrm{right}}$ have $m_s L$ and $m_r L$ columns, respectively. We also define $\mb{s} = [\mb{s}_1^T \ldots \mb{s}_{m_s}^T]^T$ and 
$\mb{r} = [\mb{r}_1^T \ldots \mb{r}_{m_r}^T]^T$. As $H \mb{x}=\mb{0}$, we have that 
\begin{equation}
\label{eq: left right}
H_{\textrm{left}} \mb{s} = H_{\textrm{right}} \mb{r}.
\end{equation}
As we consider a block BEC, some transmissions are perfect. As in App. \ref{sec: Proof of Prop. div vs Rn}, consider the case that $e_1$ blocks of length $2L$ and $e_2$ blocks of length $L$ have been erased, where $e=e_1+e_2 = |\mathcal{E}|$ is the total number of erasures, $e_1 \leq m_s$ and $e_2 \leq m_r-m_s$. Considering the structure of $H$ from (\ref{eq: general parity-check matrix form}) containing the block-diagonal matrix $H_c$, it follows that the $e_1 2L+e_2 L$ erased bits appear in only $(2e_1+e_2)(L-K) + m_r K$ of the available $(m_s +m_r)L-m_s K$ parity equations, i.e., $(2e_1+e_2)(L-K)$ equations involving $H_c$ and all $m_rK$ equations involving $H_\GLNC$. Next, $(e_1+e_2) K$ from the $m_rK$ equations involving $H_\GLNC$ cannot be used to solve erased bits in $\mb{s}$ as these equations always have at least two unknowns. The overall set of equations to decode $\mb{s}$ thus becomes
\begin{equation}
\label{eq: useful rows 2}
\left\{
\begin{array}{lr}
\mb{s}_{u_s} = \mb{y}_{u_s}^\prime & \forall ~ u_s \in \bar{\mathcal{E}} \\
H_p \mb{y}_{u_s}^\prime = \mb{0} & \forall~ u_s \in \mathcal{E} \\
\bigoplus_{{u_s} \in \mathcal{T}({u_r})} H_{u_s}^{u_r} \mb{s}_{u_s} = H_{u_r} \mb{y}_{m_s + {u_r}}^\prime & \forall~ u_r \in \bar{\mathcal{E}},
\end{array}
\right. 
\end{equation}
or, using the notation from (\ref{eq: matrix with L equations}),
\begin{equation}
\label{eq: useful rows 3}
\left\{
\begin{array}{lr}
\mb{s}_{u_s} = \mb{y}_{u_s}^\prime & \forall ~ u_s \in \bar{\mathcal{E}} \\
\bigoplus_{{u_s} \in \mathcal{T}({u_r})} \mathcal{H}_{u_s}^{u_r} \mb{s}_{u_s} = \mathcal{H}_{u_r} \mb{y}_{m_s + {u_r}}^\prime & \forall~ u_r \in \bar{\mathcal{E}},
\end{array}
\right. 
\end{equation}
where $\mb{y}_i^\prime = \frac{1+\mb{y}_i}{2}$ (BPSK modulation). We can stack the coefficients of all elements in $\mb{s}$ in a matrix $H_s$. For example, if $m_s=m_r=3$, $\mathcal{E}=\{1\}$, $\T(2)=\{1,3\}$ and $\T(3)=\{1,2\}$, then 
\begin{equation}
\label{eq: H_s}
\scriptsize
H_{s} = 
\stackrel{
\left.
\begin{array}{x{0.45cm}x{0.45cm}x{0.45cm}}
$\mb{s}_1$ & $\mb{s}_2$ & $\mb{s}_3$
\end{array}
\right.
}{
\left[ \begin{array}{x{0.45cm}x{0.45cm}x{0.45cm}}
0 & $\I$ & 0 \tn
0 & 0 & $\I$ \tn
$\mathcal{H}_{1}^{2}$ & 0 & $\mathcal{H}_{3}^{2}$ \tn
$\mathcal{H}_{1}^{3}$ & $\mathcal{H}_{2}^{3}$ & 0
\end{array}  \right]}
\normalsize
\end{equation} 
It is now easy to see that $M_{\mathcal{E}}$, as defined in Sec. \ref{sec: Coding Matrix Rank Criterion}, is closely related to $H_s$: $[M_{\mathcal{E}}]_{i,j}=1$ if $[H_s]_{(i-1)L+1 \ldots i L, (j-1)L+1 \ldots j L} \neq \mb{0}$ and $[M_{\mathcal{E}}]_{i,j}=0$ if $[H_s]_{(i-1)L+1 \ldots i L, (j-1) L+1 \ldots j L} = \mb{0}$. 

If $|\mathcal{E}| \leq d_M-1$, then $M_{\mathcal{E}}$ has full rank, according to Def. \ref{def: d_M}. As established in Lemma \ref{lemma: BP}, the set of equations represented by $M_{\mathcal{E}}$ can be solved using backward substitution. This means that at each iteration, there is an equation with only one unknown. Consider a particular iteration and denote the index of the unknown by $u$. In $H_s$, this corresponds to an equation with an unknown source-codeword vector $\mb{s}_u$ of the type 
\begin{equation}
\label{eq: useful rows 4}
\left\{
\begin{array}{l}
H_p \mb{s}_{u} = \mb{0} \\
H_{u}^{u_r} \mb{s}_{u} = \bigoplus_{\substack{u_s \in \mathcal{T}({u_r})\\ u_s \neq u}} H_{u_s}^{u_r} \mb{s}_{u_s} + H_{u_r} \mb{y}_{m_s + {u_r}}^\prime.
\end{array}
\right. 
\end{equation}
or of the type $\mb{s}_{u}=\mb{y}_{u}^\prime$. 

Under ML decoding, we obtain what was claimed if the matrices $\mathcal{H}_{u_s}^{u_r}$, ${u_s} \in \T(u_r), u_r \in \{1, \ldots, m_r\}$ have full rank. Under BP decoding, we obtain what was claimed if, for each ${u_r}$, the set of $L$ equations (\ref{eq: useful rows 4}) can be solved with BP in the case of only one unknown source-codeword vector $\mb{s}_{u}$.

\subsection{Proof of Lemma \ref{lemma: interuser failure general}}
\label{app: proof of lemma interuser}


A relay may not succeed in successfully decoding the message from a source, denoted as a failure. There are $m^2-m$ interuser channels, which all have a probability of failure. Hence, there exist $\sum_{i=0}^{m^2-m} \left( \begin{array}{c} m^2-m \\ i \end{array} \right)$ different cases, where each case corresponds to a combination of failures and successes. We denote the case where all interuser channels are successful as case 1.

Using Bayes' law, the error rate can be split:
\begin{equation}
\label{eq: bayes}
	P_{\ew} = \sum_{i} P(\textrm{case }i) P(\ew|\textrm{case }i).
\end{equation}
Defining the diversity order corresponding to each case as $d_{c,i} = -\lim_{\gamma \rightarrow \infty} \frac{\log P(\textrm{case }i) P(\ew|\textrm{case }i)}{\log \gamma}$, it follows that the overall diversity order is $d = \min_i d_{c,i}$.

The probability of $f$ failures on independent interuser channels is proportional to $\frac{1}{\gamma^f}$ \cite{tse2005fwc}, so that for this case $i$,
\begin{align}
d_{c,i} &= -\lim_{\gamma \rightarrow \infty} \frac{\log P(\textrm{case }i)}{\log \gamma} -\lim_{\gamma \rightarrow \infty} \frac{\log P(\ew|\textrm{case }i)}{\log \gamma}\\
&= f -\lim_{\gamma \rightarrow \infty} \frac{ P(\ew|\textrm{case }i)}{\log \gamma} \label{eq: d_i2parts}
\end{align}
The diversity order in the case of perfect interuser channels ($f = 0$) is $d_{c,1}$. That is, the error-correcting code can bear $d_{c,1} - 1$ erasures on node-destination links. Hence, $d_{c,i} \geq d_{c,1}$ only if $P(\ew|\textrm{case }i) \leq \frac{c}{\gamma^{d_{c,1} - f}}$, or, all information can still be retrieved at the destination, given that $f$ interuser channels and $d_{c,1} - f - 1$ node-destination channels are erased. Let us check whether this is true for all $f$.

A relay stays silent if it cannot decode all source codewords corresponding to its transmission set. If there are $f$ interuser failures, there are at most $f$ relays which stay silent in the relay phase. This corresponds to at most $f$ additional node-destination erasures adding to the assumed $d_{c,1} - f - 1$ already erased node-destination channels, yielding a total of at most $d_{c,1}-1$ erased node-destination channels, which can be supported by the code, by the definition of $d_{c,1}$.

\subsection{Proof of Lemma \ref{lemma: one failure}}
\label{app: proof of lemma interuser2}

In the case that $m_s > 4$ and Alg. \ref{alg:1} is used to construct $\{\T(u_r), u_r=1, \ldots, m_r\}$, reciprocity is irrelevant for our proposed code, as it applies that $i \notin \T(j)$ if $j \in \T(i)$. Hence, if $m_s>4$, the proof given in App. \ref{app: proof of lemma interuser} is always valid. 

Now consider the case that $d_{c,1}=2$, which corresponds to $m_s=m_r=m < 4$ (see Prop. \ref{prop: div vs Rn}). In the case of $f=1$ interuser channel, $d_{c,i}$ is always larger than one, because $P(\ew|\textrm{case }i) \leq \frac{c}{\gamma}$ as at least one channel, the source-destination channel, needs to fail to loose the corresponding information bits. 

Finally, consider the case that $m_s=m_r=m=4$ and thus $d_{c,1}=3$. Hence, in the case of no interuser failures, the code can support two node-destination failures, corresponding to four erased transmissions from two nodes, in the source phase and in the relay phase. Reciprocity is relevant as $i \in \T(j)$ if $j \in \T(i)$ for $(i,j)$ is $(1,3)$ and $(2,4)$. Because $P(\ew|\textrm{case }i) \leq \frac{c}{\gamma}$, we only have to consider the case that $f=1$, denoted as case $i$ in general.
Hence, in the case that the interuser channel between sources one and three or two and four have been erased, relays one and three or two and four, respectively, stay silent. Note that the transmission sets from the remaining active relays are disjoint when Alg. \ref{alg:1} is used, and because $n=2$, they support all sources $u_s=1, \ldots, 4$. If one node-destination channel is consequently erased, which corresponds to at most two transmissions, the destination has to recover the information bits from the erased source-codeword. Because relay $u_r$ cannot have $u_r$ in their own transmission set $\T(u_r)$, the erased relay codeword does not contain any information on the erased source-codeword, which implies that the information is in the remaining relay codeword. Hence, we have that $P(\ew|\textrm{case }i) \leq \frac{c}{\gamma^2}$ or by (\ref{eq: d_i2parts}), $d_{c,i} \geq 3$. In other words, interuser failures do not decrease the diversity order.


\small
\begin{thebibliography}{99}

\bibitem{ahl2000nif} R. Ahlswede, N. Cai, S-Y. R. Li, and R. W. Yeung,
``Network Information Flow,''
{\em IEEE Trans. on Inf. Theory}, vol.~46, no.~4, pp.~1204--1216, July 2000.

\bibitem{Bao2007gan} X. Bao and J.T. Li,
``Generalized Adaptive Network Coded Cooperation (GANCC): A Unified Framework for Network Coding and Channel Coding,''
{\em IEEE Trans. on Comm.}, vol.~59, no.~11, pp.~2934--2938, Nov. 2011.

\bibitem{bertsekas1992dn} D. Bertsekas and R. Gallager, 
``Data Networks, 2nd ed.,''
{\em Prentice Hall}, 1992.

\bibitem{biglieri1998fci} E. Biglieri, J. Proakis, and S. Shamai,
``Fading channels: information-theoretic and communications aspects,''
{\em IEEE Trans. on Inf. Theory}, vol.~44, no.~6, pp. 2619-2692, Oct. 1998.

%
%
\bibitem{bou2011dac} J.J. Boutros,
``Controlled Doping via High-Order Rootchecks in Graph Codes,''
{\em presented at IEEE Communication Theory Workshop}, Sitges, Catalonia, Spain, June 2011, \href{http://www.josephboutros.org/coding/root_LDPC_doping.pdf}{Available online} from http://www.josephboutros.org/coding/root\_LDPC\_doping.pdf.

\bibitem{bou2012tde} J.J. Boutros, D. Duyck,
``The diversity equivalence theorem and applications to codes on graphs,''
{\em in Preparation, to be submitted to IEEE Trans. on Inf. Theory}.

%

\bibitem{chou2003practical} P.A. Chou, Y. Wu, and K. Jain,
``Practical network coding,''
{\em Allerton Conf. on Communication, Control, and Computing}, Illinois, 2003.

\bibitem{cover2006eit} T.M. Cover and J.A. Thomas,
{\em Elements of Information Theory}, New York, Wiley, 2006.

\bibitem{duy2009ldg} D. Duyck, J.J. Boutros, and M. Moeneclaey,
``Low-Density Graph Codes for Slow Fading Relay Channels,''
{\em IEEE Trans. on Inf. Theory}, vol. 57, no. 7, pp.~4202--4218, July 2011.

\bibitem{duyck2010aac} D. Duyck, D. Capirone, J.J. Boutros, and M. Moeneclaey,
``Analysis and construction of full-diversity joint network-LDPC codes for cooperative communications,''
{\em Eur. Journal on Wireless Comm. and Netw.}, vol. 2010, Art. ID 805216, 2010. 

\bibitem{duyck2009afj} D. Duyck, D. Capirone, J.J. Boutros, and M. Moeneclaey,
``A full-diversity joint network-channel code construction for cooperative communications,''
{\em in Proc. IEEE Intern. Symp. on Personal, Indoor and Mob. Radio Comm.}, Tokyo, Japan, Sept. 2009. 

\bibitem{duyck2011tfd} D. Duyck, D. Capirone, M. Heindlmaier, and M. Moeneclaey,
``Towards full-diversity joint network-channel coding for large networks,''
{\em In Proc. of Europ. Wirel. Conf.}, Vienna, Austria, April 2011.

\bibitem{fab2007cmi} A. Guill\'en i F\`abregas and G. Caire,
``Coded modulation in the block-fading channel: coding theorems and code construction,''
{\em IEEE Trans. on Inf. Theory}, vol.~52, no.~1, pp. 91-114, Jan. 2006.

\bibitem{fabregas2006cit} A. Guill\'en i F\`abregas,
``Coding in the Block-Erasure Channel,''
{\em IEEE Trans. on Inf. Theory}, vol.~52, no.~11, pp.~5116--5121, Nov.~2006.

\bibitem{guo2009apj} Z. Guo, J. Huang, B. Wang, J.H. Cui, S. Zhou, and P. Willett,
``A practical joint network-channel coding scheme for reliable communication in wireless networks,''
{\em in Proc. of the ACM intern. symp. on mob. ad hoc netw. and comp.}, pp. 279-288, New Orleans, Louisiana, USA, May 2009.

\bibitem{hamdani2007cos} D. Hamdani, E. Safrianti,
``Construction of Short-Length High-Rates LDPC Codes using Difference Families,''
{\em Makara, Teknologi}, vol.~11, no.~1, pp. 25--29, Apr. 2007.

\bibitem{hausl2006jnc} C. Hausl and P. Dupraz, 
``Joint Network-Channel Coding for the Multiple-Access Relay Channel,''
{\em in Proc. IEEE Comm. Soc. on Sensor and Ad Hoc Comm. and Netw.}, vol.~3, pp.~817--822, Sept. 2006.

\bibitem{hausl2005ina} C. Hausl, F. Schreckenbach, I. Oikonomidis, and G. Bauch,
``Iterative network and channel decoding on a tanner graph,''
{\em in Proc. Allerton Conf. on Commun., Control and Computing}, Monticello, Illinois, Oct. 2005.

\bibitem{hausl2006ina} C. Hausl and J. Hagenauer, 
``Iterative Network and Channel Decoding for the Two-Way Relay Channel,''
{\em in Proc. IEEE Intern. Conf. on Comm.}, vol.~4, pp.~1568--1573, June 2006.

\bibitem{hausl2009jnc} C. Hausl,
``Joint network-channel coding for the multiple-access relay channel based on curbo codes,''
{\em Eur. Trans. Telecomms.}, vol.~20, no.~2, pp.~175--181, 2009.

\bibitem{ho2006random} T. Ho, M. M{\'e}dard, R. Koetter, D.R. Karger, M. Effros, J. Shi, and B. Leong,
``A random linear network coding approach to multicast,''
{\em IEEE Trans. on Inf. Theory}, vol.~52, no.~10, pp.~4413--4430, Oct. 2006.

\bibitem{hunter2004cc} T.E. Hunter,
{\em Coded cooperation: a new framework for user cooperation in wireless systems}, Ph.D. thesis, University of Texas at Dallas, 2004.

\bibitem{koetter2003algebraic} R. Koetter and M. M{\'e}dard,
``An algebraic approach to network coding,''
{\em In IEEE/ACM Trans. on Netw.}, vol.~11, no.~5, pp.~782--795, Oct. 2003.

\bibitem{knopp2000cbf} R. Knopp and P.A. Humblet,
``On coding for block fading channels,''
{\em IEEE Trans. on Inf. Theory}, vol.~46, no.~1, pp.~189--205, Jan. 2000.

\bibitem{kramer2005csa} G. Kramer, M. Gastpar, and P. Gupta,
``Cooperative Strategies and Capacity Theorems for Relay Networks,''
{\em IEEE Trans. on Inf. Theory}, vol.~51, no.~9, pp.~3037--3063, Sept. 2005.

\bibitem{kschischang2001fga} F. Kschischang, B. Frey, and H.-A. Loeliger,
``Factor graphs and the sum-product algorithm,''
{\em IEEE Trans. on Inf. Theory}, vol.~47, no.~2, pp. 498-519, Feb. 2001.

\bibitem{laneman2004cdw} J.N. Laneman, D. Tse, and G.W. Wornell,
``Cooperative diversity in wireless networks: Efficient protocols and outage behavior,''
{\em IEEE Trans. on Inf. Theory}, vol.~50, no.~12, pp. 3062-3080, Dec. 2004.
%
%
\bibitem{li2003lnc} S.Y.R. Li, R.W. Yeung and N. Cai,
``Linear network coding,''
{\em IEEE Trans. on Inf. Theory}, vol.~49, no.~2, pp.~371--381, Feb. 2003.

\bibitem{Li2011ncl} J. Li, J. Yuan, R. Malaney, M.H. Azmi, M. Xiao,
``Network Coded LDPC Code Design for a Multi-Source Relaying System,''
{\em IEEE Trans. on Wir. Comm.}, vol.~10, no.~5, pp.~1538--1551, May 2011.

\bibitem{Li2011bfn} J. Li, J. Yuan, R. Malaney, M. Xiao,
``Binary Field Network Coding Design for Multiple-Source Multiple-Relay Networks,''
{\em IEEE Intern. Conf. on Comm.}, June 2011.

\bibitem{malkamaki1999epc}  E. Malkamaki and H. Leib,
``Evaluating the performance of convolutional codes over block fading channels,''
{\em IEEE Trans. on Inf. Theory}, vol.~45, no.~5, pp.~1643--1646, July 1999.

\bibitem{McEliece1998tda} R.J. McEliece, D.J.C. MacKay, J.-F. Cheng,
``Turbo decoding as an instance of Pearl's ``belief propagation'' algorithm,'' 
{\em IEEE J. on Sel. Ar. in Comm.}, vol.~16, no.~2, pp.~140-152, Feb. 1998.

\bibitem{ozarow1994itc} L.H. Ozarow, S. Shamai and A.D. Wyner,
``Information theoretic considerations for cellular mobile radio,''
{\em IEEE Trans. on Veh. Techn.}, vol.~43, no.~2, pp. 359-379, May 1994.


%
%
%

\bibitem{rebellato2010muc} J.K. Rebelatto, B.F. Uch\^oa-Filho, Y. Li, and B. Vucetic, 
``Multi-User Cooperative Diversity through Network Coding Based on Classical Coding Theory,'' 
{\em IEEE Trans. on Sign. Proc.}, vol.~60, no.~2, pp.~916--926, Feb. 2012.

\bibitem{richardson2008mct} T.J. Richardson and R.L. Urbanke,
{\em Modern Coding Theory}, Cambridge University Press, 2008.

\bibitem{shakkottai2003cld} S. Shakkottai, T.S. Rappaport, and P.C. Karlsson, 
``Cross-layer design for wireless networks,''
{\em IEEE Comm. Magaz.}, vol.~41, no.~10, pp.~74--80, Oct. 2003.


\bibitem{srivastava2005cld} V. Srivastava and M. Motani, 
``Cross-layer design: a survey and the road ahead,''
{\em IEEE Communications Magazine}, vol.~43, no.~12, pp.~112--119, Dec. 2005.

%

\bibitem{tanner1981ara} M. Tanner, 
``A recursive approach to low complexity codes,''
{\em IEEE Trans. on Inf. Theory}, vol.~27, no.~5, pp. 533-547, Sept. 1981.

\bibitem{tse2005fwc} D.N.C. Tse and P. Viswanath,
{\em Fundamentals of Wireless Communication}, Cambridge University Press, May 2005.

\bibitem{ungerboeck1982ccm} G. Ungerboeck,
``Channel coding with multilevel/phase signals,''
{\em IEEE Trans. on Inf. Theory}, vol. IT-28, no. 1, pp. 55-67, 1982.

\bibitem{wymeersch2007ird} H. Wymeersch, 
{\em Iterative Receiver Design}, Cambridge University Press, 2007.

\bibitem{xiao2009muc} M. Xiao, M. Skoglund, 
``M-User Cooperative Wireless Communications Based on Non-binary Network Codes,''
{\em In Proc. Inf. Theory Workshop (ITW)}, Volos, Greece, June 2009.




\end{thebibliography}
\end{document}